\documentclass{aa}

\usepackage{graphicx}
\usepackage{xcolor}
\usepackage{txfonts}
\usepackage{hyperref}

\begin{document}

\title{The accreted Galaxy: An overview of TESS metal-poor accreted stars candidates}

   \author{Danielle de Brito Silva \inst{1,2}
          \and
         Paula Jofr\'e \inst{1,2}
         \and
         Clare Worley \inst{3,4}
         \and
         Keith Hawkins \inst{5}
         \and
         Payel Das \inst{6}
          }

   \institute{Instituto de Estudios Astrof\'isicos, Facultad de Ingenier\'ia y Ciencias, Univesidad Diego Portales, Santiago de Chile \\
              \email{danielle.debrito@mail.udp.cl}
         \and
             Millennium Nucleus ERIS \\
           \and
             School of Physical and Chemical Sciences – Te Kura Matu, University of Canterbury, Private Bag 4800, Christchurch 8140, New Zealand
           \and
             Institute of Astronomy, University of Cambridge, Madingley Road, Cambridge CB3 0HA, UK
           \and
             Department of Astronomy, The University of Texas at Austin, 2515 Speedway Boulevard, Austin, TX 78712, USA
          \and
             Department of Physics, University of Surrey, Stag Hill, Guildford, GU2 7XH, U
             }

\date{Accepted XXX. Received YYY; in original form ZZZ}

\abstract{The Milky Way is a mosaic of stars from different origins. In particular, metal-poor accreted star candidates offer a unique opportunity to better understand the accretion history of the Milky Way. In this work, we aim to explore the assembly history of the Milky Way by investigating accreted stars in terms of their ages, dynamical properties, and chemical abundances.  We also aim to better characterize the impact of incorporating asteroseismic information on age and chemical abundance calculations of metal-poor accreted stars for which \texttt{TESS} data is available. In this study, we conducted an in-depth examination of 30 metal-poor accreted star candidates, using TESS and Gaia data, as well as \texttt{MIKE} spectra. We find satisfactory agreement between seismic and predicted/spectroscopic surface gravity ($\log g$) values, demonstrating the reliability of spectroscopic data from our methodology. We found that while age determination is highly dependent on the $\log g$ and asteroseismic information used, the overall chemical abundance distributions are similar for different $\log g$. However, we found that calcium (Ca) abundances are more sensitive to the adopted $\log g$. Our study reveals that the majority of our stars have properties compatible to those reported for the Gaia-Sausage-Enceladus, with a minority of stars that might be associated to Splash. We found an age distribution with a median of $11.3_{-4.1}^{+1.3}$ Gyr when including asteroseismic information. As regarding some key chemical signatures we note that these stars are metal-poor ([Fe/H]) < -0.8), $\alpha$-rich ([$\alpha$]/Fe] > 0.2), copper-poor ([Cu/Fe] < 0 ) and with chemical abundances typical of accreted stars. These findings illustrate the importance of multi-dimensional analyses in unraveling the complex accretion history of the Milky Way.}


   \maketitle

\section{Introduction}

Over the last few decades, there has been growing evidence that the Milky Way was likely assembled through the accretion and merger events (e.g. \citealt{searle1978compositions,nissen2010two}). This evidence was supported by Gaia survey data \citep{GaiaCollaboration+2016b,brown2018gaia,brown2021gaia}, which is revolutionizing the investigation of the merger history of the Galaxy. Additionally to Gaia, information provided by spectroscopic surveys such as the Apache Point Observatory Galactic Evolution Experiment (APOGEE;\citealt{Abolfathi2018,holtzman2018apogee}) and Galactic Archaeology with Hermes (GALAH; \citealt{buder2020galah+}) have also contributed to unveiling the assembly of the Milky Way. Therefore, survey data are playing a key role in the exploration of the physical processes that led to the Milky Way as we know it today. 

Numerous works in the literature have found stars attributed to different progenitors, therefore formed \textit{ex-situ}, in contrast with the ones believed to be formed within the Milky Way (\textit{in-situ}). \cite{Belokurov2018} and \cite{Helmi2018} consolidated evidence for the relics of a fully disrupted galaxy lying in the midst of the inner stellar halo referred as Gaia-Sausage-Enceladus (GES). \cite{myeong2019evidence} found evidence of a separate accreted system in the Milky Way halo, which they called Sequoia, and is composed of stars with very retrograde motions. \cite{Koppelman2019} reported possible progenitor galaxies, including evidence for Thamnos, a distinct structure in the low-energy part of the halo. 
Additionally, \cite{horta2021evidence} found evidence for a potential past accretion event, which they referred to as Heracles, which is currently found in the central bulge, while \cite{kruijssen2020kraken} reported Kraken based on properties of the Milky Way simulated globular clusters, which have similar properties to Heracles. Therefore, there are likely multiple accreted systems in the Galaxy and distinguishing among all of them represent a challenge that require multi-dimensional data analysis.

In order to study the merger history of the Milky Way, there are three main ingredients necessary to understand its past. The first ingredient is chemical abundances, since this is associated to the birth place of a star. The second ingredient is the dynamics, since stars with the same origin in general can have similar dynamical properties. The third ingredient is stellar age, since different structures of the Galaxy and substructures in the stellar halo have different age signatures. 

The first ingredient, chemical abundances, plays a central role in Galactic Archaeology. By studying the chemical distribution of stars it is possible to infer the star formation history of the Milky Way using elements synthesised by different nucleosynthetic channels \citep{Freeman2002}. For example, $\alpha$ elements such as magnesium (Mg), calcium (Ca), and silicon (Si) and Fe-peak elements such as manganese (Mn) trace the contribution of Type II and Type Ia \citep{iwamoto99} supernovae respectively, which provide information about the star formation rate of the Milky Way. Type II and Type Ia supernovae are also valuable to investigate the contribution of material from the explosion of high- and low- mass stars in the chemical evolution of a galaxy. Studies such as \cite{nissen2010two,aguado20, matsuno2022high, carrillo2022detailed, nissen2024abundances, ceccarelli2024walk} calculated chemical abundances of accreted stars using high-resolution spectroscopy techniques in order to shed light about the evolution of their progenitor galaxies.

Together with chemical abundances, the dynamics of stars is another valuable ingredient with which to understand the Milky Way. From dynamics it is possible to separate and characterise the different components of the Galaxy and to understand its structure. In particular, dynamics can help to distinguish the building blocks of the Milky Way and also to select accreted star candidates as have been done using Gaia survey data since it was launched, as illustrated by studies such as \cite{
koppelman2020massive, naidu2020evidence, 
feuillet2021selecting,
buder2022galahaccreted}.

Stellar age is another very important ingredient that is required to unveil the past of the Galaxy, because by knowing stellar ages it is possible to understand the Galactic stellar formation timescales, the timescales of the Galactic merger history, and how the different stellar populations connect to the evolution of the Milky Way. In order to recover stellar ages, different techniques such as isochrone fitting and asteroseismology can be used, as illustrated in \cite{das2020ages, montalban2021chronologically, borre2022age}.

In the context of stellar ages, one survey with a great potential to aid in Galactic Archaeology studies is TESS \citep{ricker2014transiting}. The main objective of TESS is to find planets in bright stars close to the solar neighborhood. However, from the light curves provided by this survey it is possible to calculate masses and therefore ages using seismic information. 

TESS is an ongoing survey, that will observe stars in its 26 sectors. Since this survey will focus mostly on bright stars, the data obtained by it could be complemented by information from ground based telescopes. This will allow targets from TESS to be characterised also by other instruments, which again will contribute to the way we see the Galaxy, its stellar population content and stellar formation timescales.

Considering how attractive TESS data can be to determine stellar ages using seismic information, data from this survey could be of upmost importance in order to understand the accretion timescales experienced by the Milky Way. Unfortunately the bulk of TESS data is concentrated in metal-rich stars and seismic information of metal-poor objects is challenging to acquire. Due to that, a limited number of works in the literature explore accreted stars, which are typically metal-poor, in the light of seismic information (e.g. \citealt{montalban2021chronologically}). Therefore understanding the level of quality and precision of ages as well as chemical abundances using non-seismic information is fundamental when characterizing accreted stars. In this work we aim to explore this problem.

In this paper we calculated detailed chemical abundances of elements from different nucleosynthetic channels in order to characterize a sample of metal-poor accreted star candidates that were observed by TESS survey. This paper is organized as follows: in Section \ref{sec:data} we describe our data, while in Section \ref{sec:analysis} we describe the methodology used to recover atmospheric parameters, chemical abundances, dynamical properties and ages of our stars. In Section \ref{sec:results} we present  and discuss our results in the context of Galactic Archaeology. In Section \ref{sec:conclusion} we conclude and summarize our findings.

\section{Data}\label{sec:data}

\subsection{TESS data}

TESS is an all sky survey designed to find transitioning exoplanets, in particular exoplanets with periods of less than 13 days. TESS observes the sky in 26 segments and there are circular regions at the ecliptic poles where the segments overlap. In these regions, the TESS observing period is larger than 100 days, which contrasts with the regular 27.4 day observing period per segment and allows the discovery of longer period planets. The overlap region in the South is named the TESS Southern Continuous Viewing Zone (SCVZ).

In \cite{mackereth2021prospects}, the authors detected  the frequency of maximum power (hereafter $\nu_\mathrm{max}$) and the large frequency separation (hereafter $\Delta \nu$) for 6388 stars from the SCVZ sample and calculated ages for 1749 stars. We note that $\nu_\mathrm{max}$ refers to the frequency at which the peak power of the stellar oscillations occurs. Additionally, $\Delta \nu$ refers to the average frequency difference between consecutive modes of oscillation with the same spherical degree in the oscillation spectrum of a star. \cite{mackereth2021prospects} determined seismic information using three different pipelines: BHM \citep{elsworth2020layered}, A2Z \citep{mathur2010determining} and COR \citep{mosser2009detecting}. The authors also provided predicted values for $\nu_\mathrm{max}$ and $\Delta$$\nu$ based on \texttt{2MASS} photometry and Gaia DR2 parallax, using the \texttt{asteroestimate} \footnote{\url{https://github.com/jmackereth/asteroestimate}} package, which is an implementation of the formalism presented in \cite{chaplin2011predicting}, \cite{campante2016asteroseismic} and \cite{schofield2019asteroseismic}, as explained in more detail in \cite{mackereth2021prospects}. 

\subsection{Spectroscopic observations}

\begin{table*}[hbt!]
\centering
\caption{Gaia DR3 ID, ID adopted in this work, RA, DEC and G magnitude of the 30 stars explored in this work.}
\label{tab:basic_info}
\begin{tabular}{|r|l|r|r|r|r|l|}
\hline
  \multicolumn{1}{|c|}{Gaia DR3 SOURCE ID} &
  \multicolumn{1}{c|}{ID} &
  \multicolumn{1}{c|}{RA (deg)} &
  \multicolumn{1}{c|}{DEC (deg)} &
  \multicolumn{1}{c|}{G (mag)} \\
\hline
  4769356093226280832 & 401 & 81.911 & -56.785 & 10.02\\
  4778230075479363840 & 402 & 70.334 & -51.913 & 9.39\\
  5504082885695367936 & 403 & 109.066 & -51.590 & 9.33\\
  4695854977859763456 & 404 & 40.655 & -66.908 & 9.33\\
  4670830746046529920 & 405 & 53.668 & -67.234 & 10.98\\
  5268456680571547904 & 406 & 103.022 & -69.491 & 10.40\\
  5270675018297844224 & 407 & 117.785 & -69.466 & 10.56\\
  4676601464106231040 & 408 & 65.931 & -62.285 & 10.32\\
  5480342058829420800 & 409 & 103.603 & -60.031 & 10.41\\
  5302209492824299520 & 410 & 127.802 & -60.222 & 10.67\\
  5291442147248954240 & 411 & 123.622 & -58.363 & 8.52\\
  5314537492071304832 & 412 & 125.253 & -59.343 & 10.01\\
  4639772360060363520 & 413 & 48.779 & -74.720 & 10.98\\
  4627471127048415744 & 414 & 65.807 & -77.929 & 10.75\\
  4641343802694760448 & 415 & 49.805 & -73.648 & 10.20\\
  5221447924214761216 & 416 & 127.342 & -71.538 & 10.70\\
  4650709717625944704 & 417 & 89.581 & -72.516 & 10.75\\
  5267028900300303488 & 418 & 107.630 & -71.364 & 10.60\\
  4641929116541889792 & 419 & 53.698 & -72.045 & 10.42\\
  5194368911329631104 & 420 & 114.689 & -83.635 & 8.80\\
  5277009476587267328 & 421 & 127.913 & -62.745 & 9.79\\
  4648545500782055040 & 422 & 86.380 & -74.955 & 10.50\\
  5219929017623667840 & 423 & 129.444 & -73.548 & 9.32\\
  4663217074702729088 & 424 & 70.302 & -65.312 & 9.18\\
  4665485023591893888 & 425 & 73.038 & -62.153 & 9.84\\
  5477802290050446720 & 426 & 90.147 & -63.957 & 9.67\\
  5282079908816150912 & 427 & 105.532 & -65.561 & 8.50\\
  5218217662135843456 & 428 & 146.296 & -72.388 & 10.70\\
  4615097218893632256 & 429 & 58.123 & -84.991 & 10.08\\
  4765580920052236160 & 430 & 89.200 & -57.126 & 8.73\\
\hline\end{tabular}
\end{table*}

In this work, we selected stars from the catalogue of  \cite{mackereth2021prospects}, which contains 15 405 red giant stars brighter than G $\leq$ 11 that were targeted as part of the SCVZ.  Our selection consisted of choosing only the stars with retrograde motions following the methodology described in Section  \ref{subsec:velocities_dynamics_methods}, in order to select accreted stars candidates. Several works in the literature (e.g. \citealt{nissen2010two,Belokurov2018,Helmi2018,myeong2019evidence}) have shown that accreted stars tend to have retrograde motions. This  characteristic can be caused by the angle at which the progenitor galaxy was disrupted and incorporated into the Milky Way. This resulted in a sample of 30 stars. Their identification number in different catalogues (GAIA ID and the ID adopted in this work), positions and magnitudes are presented in Table \ref{tab:basic_info}. In order to do this selection, we used EDR3 data of the Gaia survey, since we observed the stars in January 2021.

We note that for the stars studied in this work there is overall a good agreement on the parallax from the different Gaia data releases. Between Gaia DR2 and Gaia DR3, for example, the mean difference between the different parallax values is -0.04 mas with a standard deviation of 0.04 mas. The larger difference observed was 0.14 mas, but all the stars except from one, have differences below 0.10 mas. There are only two stars in the sample where the two parallax values do not agree within 3 sigmas: stars  419 and star 422. However, we did not find any relevant systematic effects associated with these objects in this work.

Among the chosen 30 stars, only two of them have ages provided by \cite{mackereth2021prospects}, and 21 stars have $\nu_\mathrm{max}$ and $\Delta$$\nu$. Therefore, we used the $\nu_\mathrm{max}$ and $\Delta$$\nu$ values provided to determine our own stellar ages, as described in Section \ref{subsec:ages_methods}. We also considered the $\nu_\mathrm{max}$ to derive surface gravity using scaling relations. 

In order to determine chemical abundances  we took high-resolution (R $\approx$ 35 000) spectra using \texttt{MIKE} spectrograph on the 6.5 m Clay Telescope of Las Campanas Observatory in January 2021. \texttt{MIKE} spectrograph is a double echelle spectrograph, composed of a blue and a red arm. The blue arm provides a wavelength coverage from approximately 3350 to 5000 ${\AA}$, while the red arm provides a wavelength coverage from approximately 4900 to 9500 ${\AA}$. These spectra have a signal-to-noise ratio of 100  pixel$^{-1}$ on average. The data were reduced using the pipeline for data reduction of the instrument: the Carnegie Python \texttt{MIKE} pipeline \citep{kelson2003optimal}, hereafter CarPy. CarPy produces a science ready 2D spectrum considering flats for slit distortion tracing and flat fielding, and ThaAr arcs for wavelength solution. In case of multiple exposures of the same object, CarPy automatically co-adds the spectra of the same objects.

In order to stack the orders of the \texttt{MIKE} spectra, a step that is not performed during the data reduction procedure of Carpy, we first normalized each order using splines of 3 degrees every 5 nm using the tool for the treatment and analysis of stellar spectra \texttt{iSpec} (\citeauthor{Blanco2014} \citeyear{Blanco2014}, \citeauthor{Blanco2019} \citeyear{Blanco2019}). Then, we corrected the spectra by the radial velocity of the stars also using \texttt{iSpec}. To determine the radial velocity, we performed a cross-correlation between the observed spectrum and a spectrum of Arcturus from an Atlas provided with \texttt{iSpec}. The radial velocity was done order by order in order to avoid issues related to inaccurate wavelength calibration, as described in \citealt{de2022j01020100}. The subsequent combination of orders was done using the IRAF package \citep{iraf1993}.

Only three of seven stars have a spectrum taken by APOGEE or GALAH, which have undertaken a follow-up campaign of the TESS-SCVZ field as well. We note that even though the 30 stars chosen in our selection were mostly  not observed by spectroscopic surveys, other stars studied in \cite{mackereth2021prospects} with prograde motions are in GALAH survey. The GALAH survey is a spectroscopic survey that delivers chemical abundances of up to 30 elements and we used its chemical abundances in this work as reference to discuss our findings when compared with the prograde sample presented in \cite{mackereth2021prospects}. We also selected a sample of stars with retrograde motions in GALAH, as an additional control sample.

\section{Methods} \label{sec:analysis}

\subsection{Kinematics}\label{subsec:velocities_dynamics_methods}

Since we have Gaia DR3 data for our stellar sample, we have information about their total velocities.  We calculated the velocities $(\mathrm{U},\mathrm{V}, \mathrm{W})$ adopting Galactic radius $\mathrm{R}_\odot = 8.2$~Kpc \citep{mcmillan2016mass} for the solar position and Galactic height of $\mathrm{z}_\odot = 0.0025$~Kpc  \citep{juric2008milky} and $(\mathrm{U},\mathrm{V}, \mathrm{W})_{\odot} = (11.10, 247.97, 7.25)$~km/s for the solar velocity relative to the Galactic Center, following \cite{matsuno2020star}. We used \texttt{astropy} package \citep{robitaille2013astropy, price2018astropy} for the task. Here we used Gaia DR3 positions, proper motions, parallaxes and radial velocities. We adopted the following convention: $\mathrm{U}$ positive toward Galactic Center, $\mathrm{V}$ positive in the direction of Galactic rotation and $\mathrm{W}$ positive toward the North Galactic Pole. The final velocity in assumed to be in the Galactic rest frame. Retrograde stars are those with $\mathrm{V} < 0$.

We integrated the orbit of these stars and calculated the total energy ($\mathrm{E}$), the angular momentum in the $z$ direction ($\mathrm{L}_z$) and the eccentricities ($\mathrm{ecc}$) using the \texttt{gala} Python package \citep{price2018astropy}. We adopted in the orbit integration the default Milky Way potential and the Galactocentric coordinate frame built using the values described above.

\subsection{Stellar parameters: Effective temperature and surface gravity}\label{subsec:teff_logg_methods}

The effective temperature ($\mathrm{T}_\mathrm{eff}$) was calculated using the InfraRed Flux Method (IRFM) as presented in \cite{casagrande2021galah}.  The authors implement Gaia and \texttt{2MASS} photometry \citep{skrutskie2006} in the IRFM and apply it to stars in GALAH DR3 considering different evolutionary stages, metallicities ([Fe/H]) and surface gravities ($\log g$). In order to calculate $\mathrm{T}_\mathrm{eff}$ we applied the \texttt{colte} code \footnote{https://github.com/casaluca/colte}, the color-$\mathrm{T}_\mathrm{eff}$ relations from \cite{casagrande2021galah}, Gaia EDR3 and \texttt{2MASS}  photometry, as well as Gaia EDR3 ($\log g$) and [Fe/H]. Since the \texttt{colte} code provided every permutation of the IRFM colour combinations for the $\mathrm{T}_\mathrm{eff}$, we calculated the mean absolute deviation (MAD) of all the values provided. We then multiplied the MAD value by 1.48 in order to obtain an approximation of the standard deviation. The typical uncertainty of $\mathrm{T}_\mathrm{eff}$ calculated with this approximation is of the order of 60 K. We then considered the standard deviation as the uncertainty of $\mathrm{T}_\mathrm{eff}$.

Using our derived $\mathrm{T}_\mathrm{eff}$, and the mean $\nu_\mathrm{max}$ calculated from all the available values from \cite{mackereth2021prospects} (see Section \ref{sec:data}), we determined $\log g$ using the scaling relation \citep{kjeldsen1994amplitudes}:

\begin{equation}
    \log g = \log \nu_{\mathrm{max}} + \log g_{\odot} + \frac{1}{2} \log \mathrm{T}_\mathrm{eff} - \frac{1}{2} \log \mathrm{T}_{\mathrm{eff}, \odot} - \log \nu_{\mathrm{max}, \odot}
    \label{eq:logg_calc}
\end{equation}

Since not all stars have a value of $\nu_\mathrm{max}$ and $\Delta \nu$, we also considered other homogeneous determinations for $\log g$, namely one $\log g$ using only the predicted $\nu_\mathrm{max}$ from \cite{mackereth2021prospects} and Eq. \ref{eq:logg_calc},  and the $\log g$ derived spectroscopically. The spectroscopic $\log g$ was calculated using the \texttt{iSpec} code and their determination is better described in Section \ref{subsec:spectroscopy_methods}.

In summary, we have three different estimates of $\log g$: seismic (using the mean of $\nu_\mathrm{max}$ and Eq. \ref{eq:logg_calc}), predicted (using only the predicted $\nu_\mathrm{max}$  from \cite{mackereth2021prospects} and Eq. \ref{eq:logg_calc}), and spectroscopic (using \texttt{MIKE} spectral features). In Table \ref{tab:data_atmospheric}, we present the values of $\mathrm{T}_\mathrm{eff}$, seismic $\log g$, predicted $\log g$, spectroscopic $\log g$, [Fe/H], microturbulence velocity (v$_{\text{mic}}$), $\nu_\mathrm{max}$ and $\Delta \nu$ for the stars studied in this work. We note that  $\nu_\mathrm{max}$ and $\Delta \nu$ are the values obtained after considering the quality criteria described in Section \ref{subsec:teff_logg_results}.

\begin{table*}[hbt!]
\centering
\caption{Values of $\mathrm{T}_\mathrm{eff}$, seismic $\log g$, predicted $\log g$, spectroscopic $\log g$, [Fe/H], microturbulence velocity (v$_{\text{mic}}$), $\nu_\mathrm{max}$ and $\Delta \nu$ adopted for our sample of stars. We note that the values for [Fe/H] and v$_{\text{mic}}$ presented in this table were obtained using the spectroscopic $\log g$, which allowed it to be available for the 30 stars presented in this work. We also note that while a substantial part of our study is focused on the stars for which $\nu_\mathrm{max}$ and $\Delta \nu$ were available, in Section \ref{subsec:chemical_abundance_distribution} we present the chemical distribution of the stars showed in this table, even if seismic information was not available for the object.}
\label{tab:data_atmospheric}
\begin{tabular}{|r|r|r|r|r|r|r|r|r|}
\hline
  \multicolumn{1}{|c|}{ID} &
  \multicolumn{1}{c|}{$\mathrm{T}_\mathrm{eff}$} &
  \multicolumn{1}{c|}{seismic} &
  \multicolumn{1}{c|}{predicted} &
  \multicolumn{1}{c|}{spectroscopic} &
  \multicolumn{1}{c|}{[Fe/H]} &
  \multicolumn{1}{c|}{v$_{\text{mic}}$}&
  \multicolumn{1}{|c|}{$\nu_\mathrm{max}$} &
  \multicolumn{1}{c|}{$\Delta \nu$} \\
  \multicolumn{1}{|c|}{} &
  \multicolumn{1}{c|}{(K)} &
  \multicolumn{1}{c|}{$\log g$} &
  \multicolumn{1}{c|}{$\log g$} &
  \multicolumn{1}{c|}{$\log g$} &
  \multicolumn{1}{c|}{dex} &
  \multicolumn{1}{c|}{(km/s)} &
  \multicolumn{1}{|c|}{($\mu$Hz)} &
  \multicolumn{1}{c|}{($\mu$Hz)} \\
\hline
401 & $4337 \pm 10$ & - & $1.13 \pm 0.04$ & $1.05 \pm 0.07$ & $-1.11 \pm 0.02$ & $1.78 \pm 0.05$ & - & - \\
402 & $4697 \pm 52$ & - & $1.34 \pm 0.05$ & $1.25 \pm 0.10$ & $-1.66 \pm 0.01$ & $2.05 \pm 0.08$ & - & - \\
403 & $4476 \pm 43$ & - & $1.17 \pm 0.37$ & $1.91 \pm 0.09$ & $-1.45 \pm 0.03$ & $0.95 \pm 0.06$ & - & - \\
404 & $4923 \pm 117$ & - & $2.07 \pm 0.05$ & $1.96 \pm 0.09$ & $-1.45 \pm 0.01$ & $1.58 \pm 0.09$ & - & - \\
405 & $5297 \pm 35$ & $2.07 \pm 0.01$ & $2.31 \pm 0.01$ & $2.30 \pm 0.12$ & $-1.26 \pm 0.01$ & $1.58 \pm 0.11$ & $3.64 \pm 0.49$ & $0.68 \pm 0.12$ \\
406 & $5112 \pm 96$ & $2.29 \pm 0.01$ & $2.37 \pm 0.11$ & $2.60 \pm 0.10$ & $-0.58 \pm 0.01$ & $1.77 \pm 0.08$ & $23.47 \pm 1.50$ & $3.53 \pm 0.29$ \\
407 & $5243 \pm 100$ & $2.94 \pm 0.01$ & $2.84 \pm 0.11$ & $3.07 \pm 0.08$ & $-0.66 \pm 0.01$ & $1.45 \pm 0.09$ & $104.65 \pm 1.24$ & $9.22 \pm 0.61$ \\
408 & $5342 \pm 60$ & $3.12 \pm 0.01$ & $3.11 \pm 0.02$ & $3.06 \pm 0.10$ & $-1.30 \pm 0.01$ & $1.28 \pm 0.13$ & $155.54 \pm 3.30$ & $10.43 \pm 0.32$  \\
409 & $4854 \pm 37$ & $1.76 \pm 0.01$ & $1.82 \pm 0.06$ & $1.52 \pm 0.11$ & $-1.17 \pm 0.02$ & $1.52 \pm 0.10$ & $7.08 \pm 0.57$ & $1.21 \pm 0.50$ \\
410 & $5094 \pm 36$ & - & $2.42 \pm 0.23$ & $1.97 \pm 0.11$ & $-1.28 \pm 0.01$ & $1.23 \pm 0.11$ & - & - \\
411 & $4335 \pm 74$ & - & $0.76 \pm 0.14$ & $1.03 \pm 0.09$ & $-1.55 \pm 0.01$ & $2.01 \pm 0.05$ & - & - \\
412 & $4996 \pm 99$ & $2.76 \pm 0.01$ & $2.76 \pm 0.02$ & $2.72 \pm 0.11$ & $-0.64 \pm 0.01$ & $1.36 \pm 0.08$ & $69.90 \pm 1.68$ & $7.52 \pm 0.07$ \\
413 & $4850 \pm 85$ & $1.97 \pm 0.01$ & $2.00 \pm 0.03$ & $2.05 \pm 0.09$ & $-1.09 \pm 0.01$ & $1.54 \pm 0.08$ & $11.50 \pm 1.12$ & $1.68 \pm 0.19$ \\
414 & $4598 \pm 71$ & $1.87 \pm 0.02$ & $1.76 \pm 0.16$ & $2.07 \pm 0.09$ & $-0.73 \pm 0.01$ & $1.58 \pm 0.08$ & $8.88 \pm 0.16$ & $1.45 \pm 0.24$ \\
415 & $4342 \pm 16$ & $1.79 \pm 0.01$ & $1.73 \pm 0.14$ & $2.01 \pm 0.07$ & $-0.39 \pm 0.01$ & $1.65 \pm 0.07$ & $7.97 \pm 1.36$ & $1.27 \pm 0.01$  \\
416 & $4600 \pm 27$ & - & $1.63 \pm 0.21$ & $2.06 \pm 0.10$ & $-0.33 \pm 0.01$ & $1.79 \pm 0.07$ & - & - \\
417 & $5314 \pm 84$ & - & $3.56 \pm 0.04$ & $3.65 \pm 0.08$ & $-0.63 \pm 0.01$ & $1.24 \pm 0.10$ & - & -  \\
418 & $5211 \pm 55$ & $2.70 \pm 0.01$ & $2.74 \pm 0.04$ & $2.65 \pm 0.09$ & $-1.29 \pm 0.01$ & $1.66 \pm 0.10$ & $59.81 \pm 3.12$ & $6.61 \pm 0.06$ \\
419 & $4658 \pm 36$ & $1.50 \pm 0.01$ & $1.47 \pm 0.03$ & $1.53 \pm 0.09$ & $-1.40 \pm 0.02$ & $1.48 \pm 0.06$ & $3.92 \pm 0.35$ & $0.76 \pm 0.02$ \\
420 & $5398 \pm 47$ & - & $2.55 \pm 0.20$ & $2.14 \pm 0.09$ & $-1.02 \pm 0.01$ & $2.10 \pm 0.10$ & - & - \\
421 & $4802 \pm 32$ & -  & $1.50 \pm 0.04$ & $1.59 \pm 0.08$ & $-1.18 \pm 0.01$ & $1.76 \pm 0.06$ & -  & - \\
422 & $4956 \pm 84$ & $2.52 \pm 0.01$ & $2.45 \pm 0.14$ & $2.76 \pm 0.09$ & $-0.61 \pm 0.03$ & $1.49 \pm 0.07$ & $40.91 \pm 0.24$ & $4.86 \pm 0.16$ \\
423 & $4744 \pm 99$ & - & $1.76 \pm 0.06$ & $1.87 \pm 0.14$ & $-1.13 \pm 0.01$ & $1.67 \pm 0.06$ & - & - \\
424 & $5406 \pm 58$ & $2.89 \pm 0.01$ & $2.71 \pm 0.05$ & $2.61 \pm 0.13$ & $-1.67 \pm 0.04$ & $1.71 \pm 0.09$ & $91.17 \pm 0.19$ & $8.31 \pm 0.05$ \\
425 & $4865 \pm 50$ & $1.70 \pm 0.01$ & $1.69 \pm 0.04$ & $1.77 \pm 0.17$ & $-2.32 \pm 0.07$ & $1.82 \pm 0.09$ & $6.14 \pm 0.09$ & $0.99 \pm 0.01$ \\
426 & $4609 \pm 75$ & $1.65 \pm 0.01$ & $1.65 \pm 0.05$ & $1.55 \pm 0.11$ & $-0.94 \pm 0.01$ & $1.67 \pm 0.06$ & $5.68 \pm 0.06$ & $1.13 \pm 0.32$ \\
427 & $4285 \pm 30$ & - & $0.65 \pm 0.07$ & $0.78 \pm 0.08$ & $-1.62 \pm 0.02$ & $2.13 \pm 0.05$ & - & - \\
428 & $4156 \pm 78$ & - & $1.08 \pm 0.04$ & $1.17 \pm 0.14$ & $-0.82 \pm 0.01$ & $1.80 \pm 0.06$ & - & -  \\
429 & $4649 \pm 53$ & - & $1.25 \pm 0.07$ & $1.40 \pm 0.10$ & $-1.59 \pm 0.05$ & $1.00 \pm 0.09$ & - & - \\
430 & $4336 \pm 67$ & $1.98 \pm 0.02$ & $0.99 \pm 0.00$ & $1.00 \pm 0.09$ & $-1.30 \pm 0.02$ & $2.24 \pm 0.07$ & $22.77 \pm 1.54$ & $2.65 \pm 0.37$  \\
\hline
\end{tabular} 
\end{table*}

\subsection{Spectroscopy}\label{subsec:spectroscopy_methods}

In order to calculate the chemical abundances we followed the procedure presented in \cite{de2022j01020100} who determined parameters and abundances of an apparently high-mass accreted star candidate that might be the product of binary evolution using data from \texttt{MIKE} spectrograph.

In short, we used \texttt{iSpec} for the chemical abundance determination. This method consists of fitting the chosen spectral regions on-the-fly until the synthetic spectra agrees with the observed one. We used the spectral synthesis method for which we adopted the line list from the Gaia-ESO survey \citep{heiter2021atomic}, the solar abundances from \cite{grevesse2007},
the one-dimensional atmospheric models \texttt{MARCS7} \citep{gustafsson2008} and the radiative transfer code \texttt{TURBOSPECTRUM} \citep{gray1994}, which assumes Local Thermodynamical Equilibrium (LTE).

With the aim of determining the spectroscopic $\log g$ we also used iSpec. We fixed the $\mathrm{T}_\mathrm{eff}$ at the value obtained using the IRFM,  $v \sin i$ at 1.6 km/s, following \cite{Blanco2019}, and $v_\mathrm{mac}$ and $v_\mathrm{mic}$ were determined from the same empirical relation used in Gaia-ESO. We let the metallicity be a free parameter. Only Fe I and Fe II lines were used  to determine the spectroscopic $\log g$, applying the method of ionisation and excitation balance \citep{gray2005observation}. We note that
when the predicted or the seismic $\log g$ were used, the metallicity was the only parameter let free.  

\begin{figure}
\centering
\includegraphics[width=8.5cm]{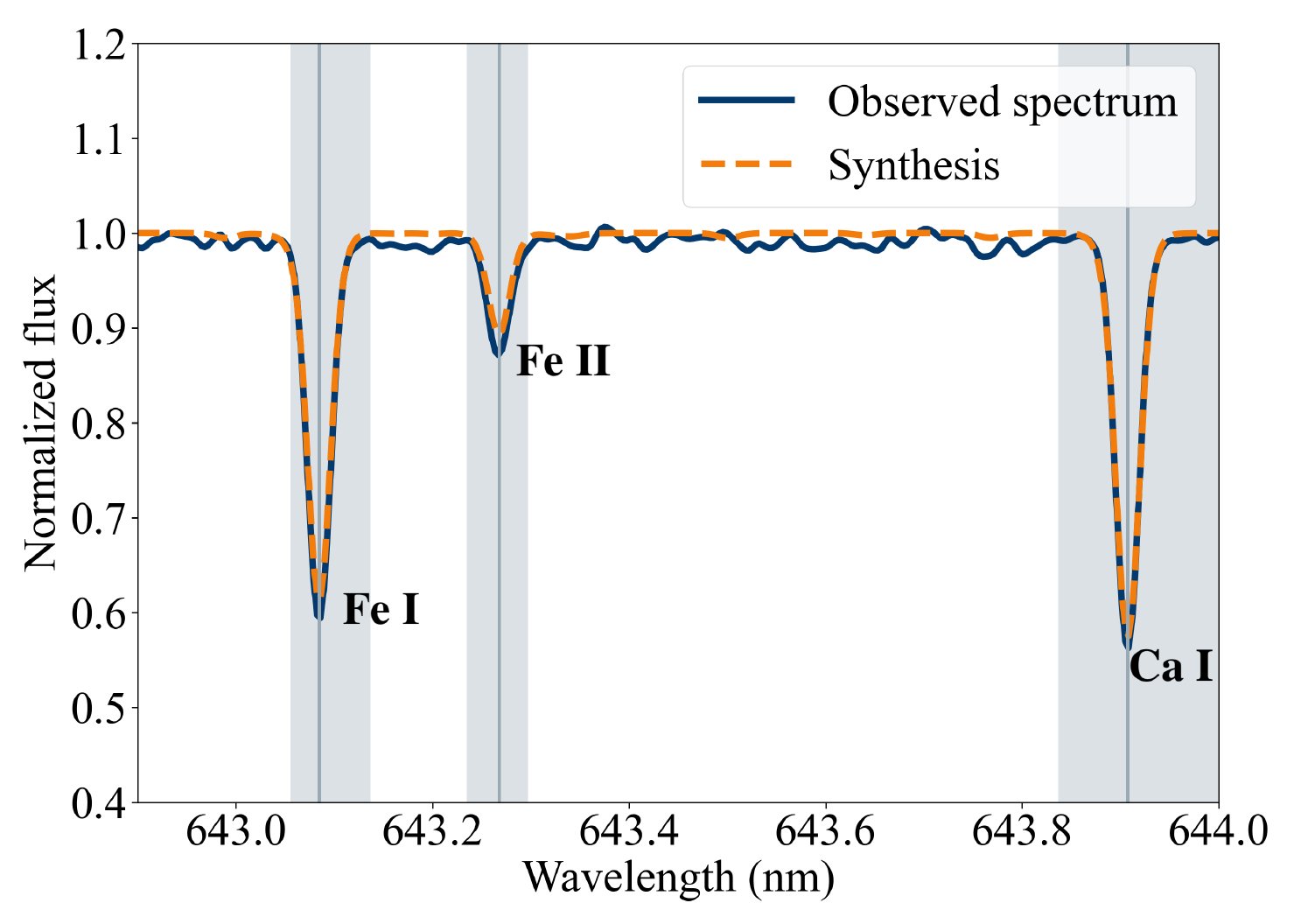}
\includegraphics[width=8.5cm]{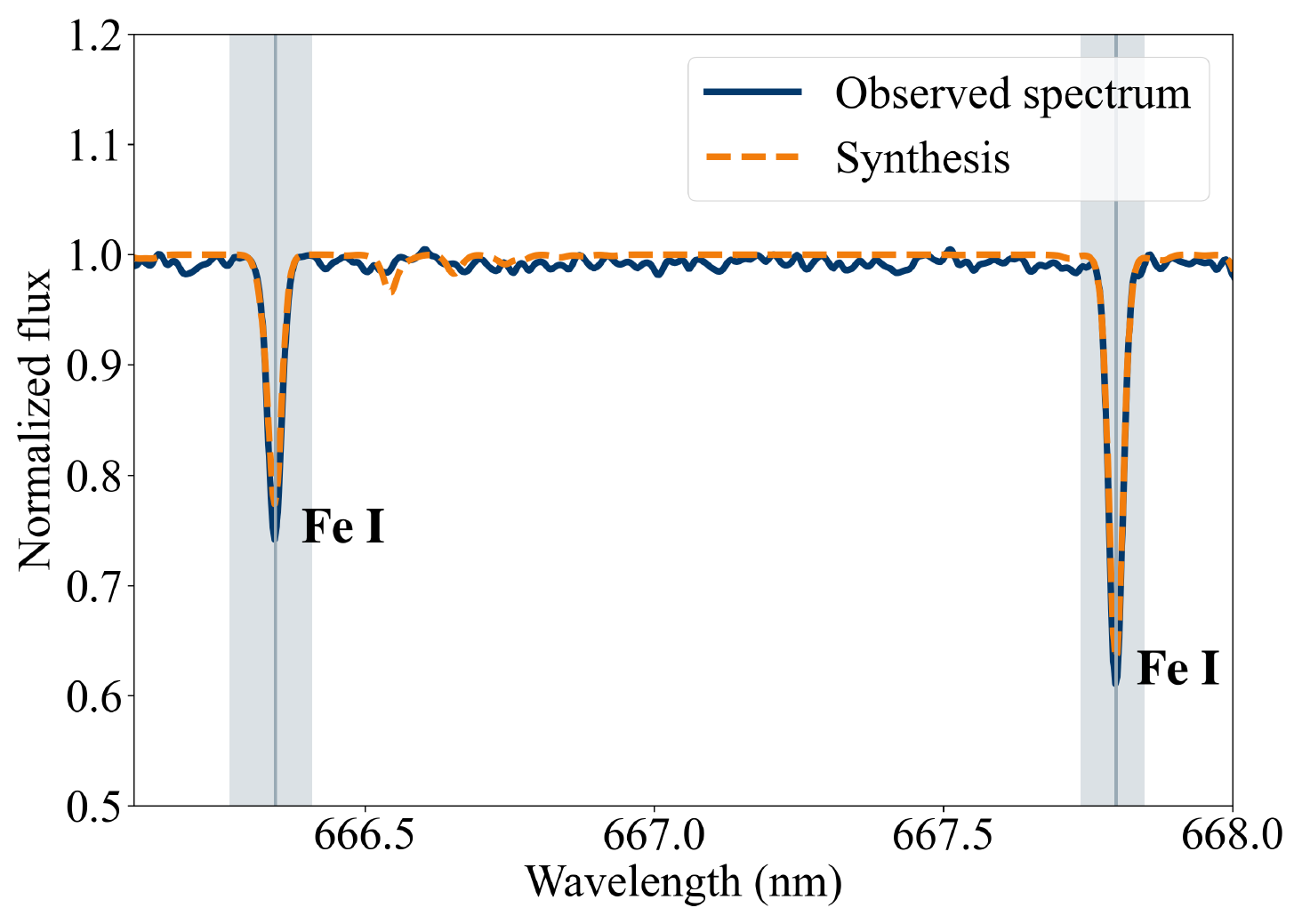}\\
\caption{{Example region of stellar spectra. In blue we show an example of spectrum obtained with \texttt{MIKE} spectrograph at Clay Telescope at Las Campanas Observatory. In orange we show a synthetic spectrum built using \texttt{iSpec} \citep{Blanco2014,Blanco2019}. The gray regions were used to fit the observed spectra to obtain chemical abundances of different elements.}}
\label{fig:spectrum}
\end{figure}

The chemical abundances were determined line-by-line following the lines used in \cite{de2022j01020100}, and low quality results (i.e. chemical abundances with individual line uncertainties higher than 0.40 dex) were discarded. We opted to analyse the different exposures of the spectra separately, in order to obtain more measurements and assess our precision. We also evaluated the quality of our results by visually inspecting how well the synthetic spectrum represented the observed one. One example of the fit obtained is presented in Figure \ref{fig:spectrum}. In this figure we show in blue the observed spectrum, in orange the synthetic spectrum, and the gray areas are the regions considered for the fit. The vertical lines indicate the center of the lines fitted inside each region.
Our method was applied to the \texttt{MIKE} spectra which allowed us to determine metallicities and chemical abundances of 15 elements (Ti, Ni, Ca, Si, Mg, Sc, Cr, Co, Mn, Ba, V, Y, Cu, Na and Al). 

\subsection{Age determination}\label{subsec:ages_methods}

We determined the stellar ages using the BAyesian STellar Algorithm (\texttt{BASTA}, \citealt{silva2015ages, aguirre2022bayesian}) \footnote{\url{http://stev.oapd.inaf.it/cgi-bin/param}}. \texttt{BASTA} provides stellar properties by comparing a grid of stellar models with observational data. We calculated three stellar ages with \texttt{BASTA}, based on the three different $\log g$ we adopted in this work:

\begin{itemize}
    \item Seismic case: we provided \texttt{BASTA} as input $\mathrm{T}_\mathrm{eff}$, metallicity, [$\alpha$/Fe], parallax, \texttt{2MASS} photometry, $\Delta \nu$, and $\nu_\mathrm{max}$.
    \item Predicted case: we used as input to \texttt{BASTA} $\mathrm{T}_\mathrm{eff}$, metallicity, [$\alpha$/Fe], parallax, \texttt{2MASS} photometry, and predicted $\log g$.
    \item Spectroscopic case: we used as input $\mathrm{T}_\mathrm{eff}$, metallicity, [$\alpha$/Fe], parallax, \texttt{2MASS} photometry, and spectroscopic $\log g$.
\end{itemize}

The \texttt{2MASS} photometry includes the magnitudes of the bands \texttt{J}, \texttt{H}, \texttt{K$_{s}$} corrected for extinction using the dust map presented in \cite{Schlegel1998}. We note that for the age calculation, [$\alpha$/Fe] was approximated to be equal to [Mg/Fe].

Within \texttt{BASTA}, we used the updated Bag of Stellar Tracks and Isochrones (\texttt{BASTI}, \citealt{hidalgo2018updated,pietrinferni2021updated}) grid, which cover masses from 0.1 to 15 M$_{\sun}$, [Fe/H] from -3.2 to 0.45 dex, and ages from 20 Myr to 14.5 Gyrs. We defined an upper limit for the ages of 13.7 Gyr, in order to avoid age solutions older than the age of the universe. Following \cite{borre2022age}, which calculated ages using \texttt{BASTA} for 23 \textit{ex-situ} stars using Kepler \citep{borucki2010kepler,koch2010kepler} and K2 \citep{howell2014k2} seismic data, we adopted the Salpeter initial mass function \citep{salpeter1955luminosity}, in order to favor the presence of low-mass stars, since there are evidence that accreted stars are old objects (e.g. \citealt{das2020ages,montalban2021chronologically,borre2022age}). Finally, we configured \texttt{BASTA} to account both for mass-loss and diffusion. \texttt{BASTA} follows the mass-loss prescription described in \cite{reimers1975circumstellar} and the diffusion described in \cite{thoul1993element}.

\section{Results and discussion} \label{sec:results}

\subsection{Dynamics}\label{subsec:dynamics_results}

Studies in the literature have identified accreted stars within the L$_{z}$-E plane, associating different regions with various building blocks of the Milky Way. For instance, \cite{Koppelman2019} attributed regions in this diagram to stars originating from possible progenitor galaxies such as GES, Sequoia, Helmi Streams \citep{Helmi1999}, Thamnos 1 and Thamnos 2. Additionally, \cite{naidu2020evidence} found 4 new substructures in the halo, possibly associated with different progenitor galaxies: Aleph, Arjuna, I’itoi and Wukong. Furthermore, \cite{horta2022chemical} demonstrated the presence of Galactic substructures likely associated with different progenitor galaxies in the L$_{z}$-E plane, including regions likely linked to GES, Sequoia, Heracles, among others.

\begin{figure}
\centering
\includegraphics[width=8cm]{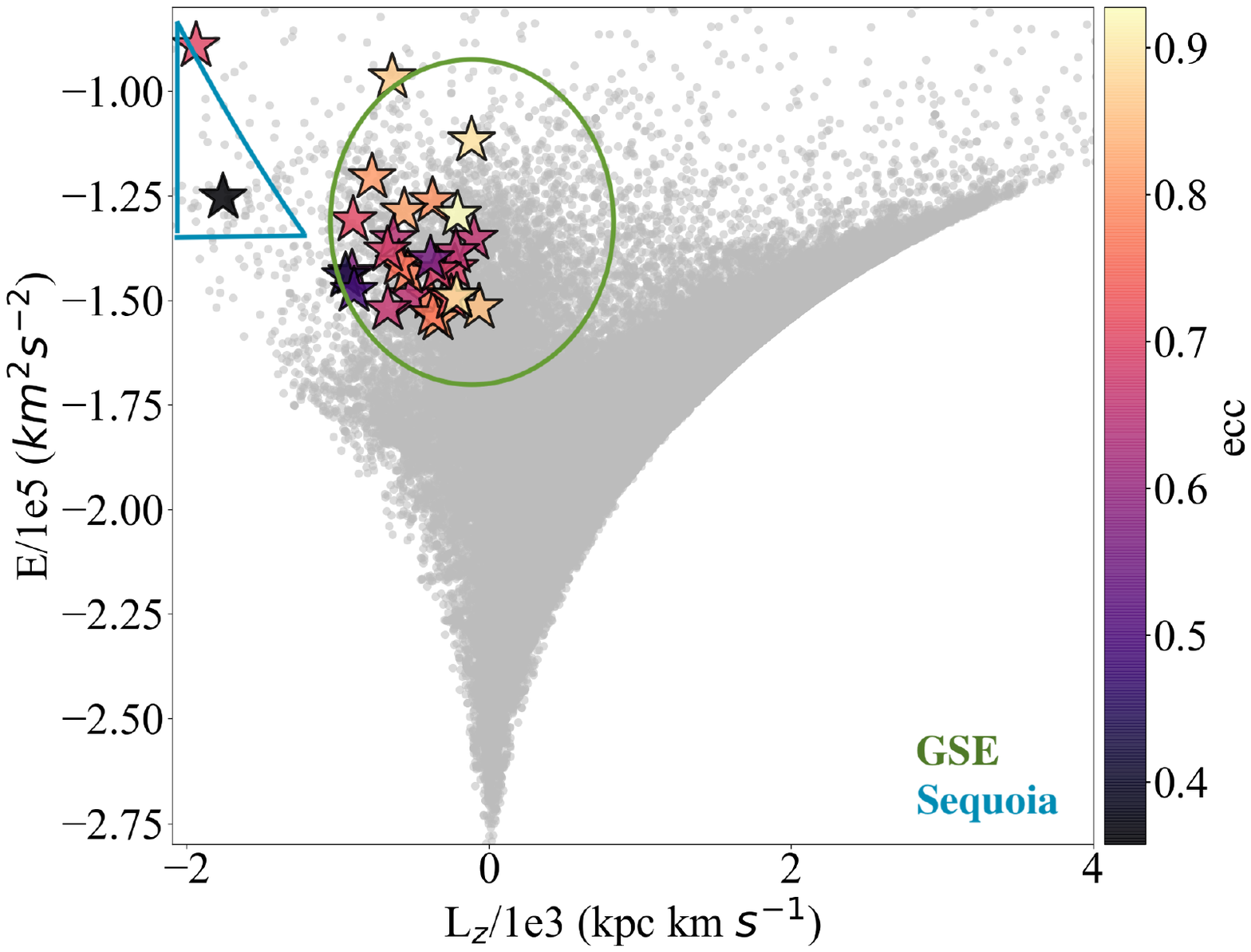}
\includegraphics[width=8cm]{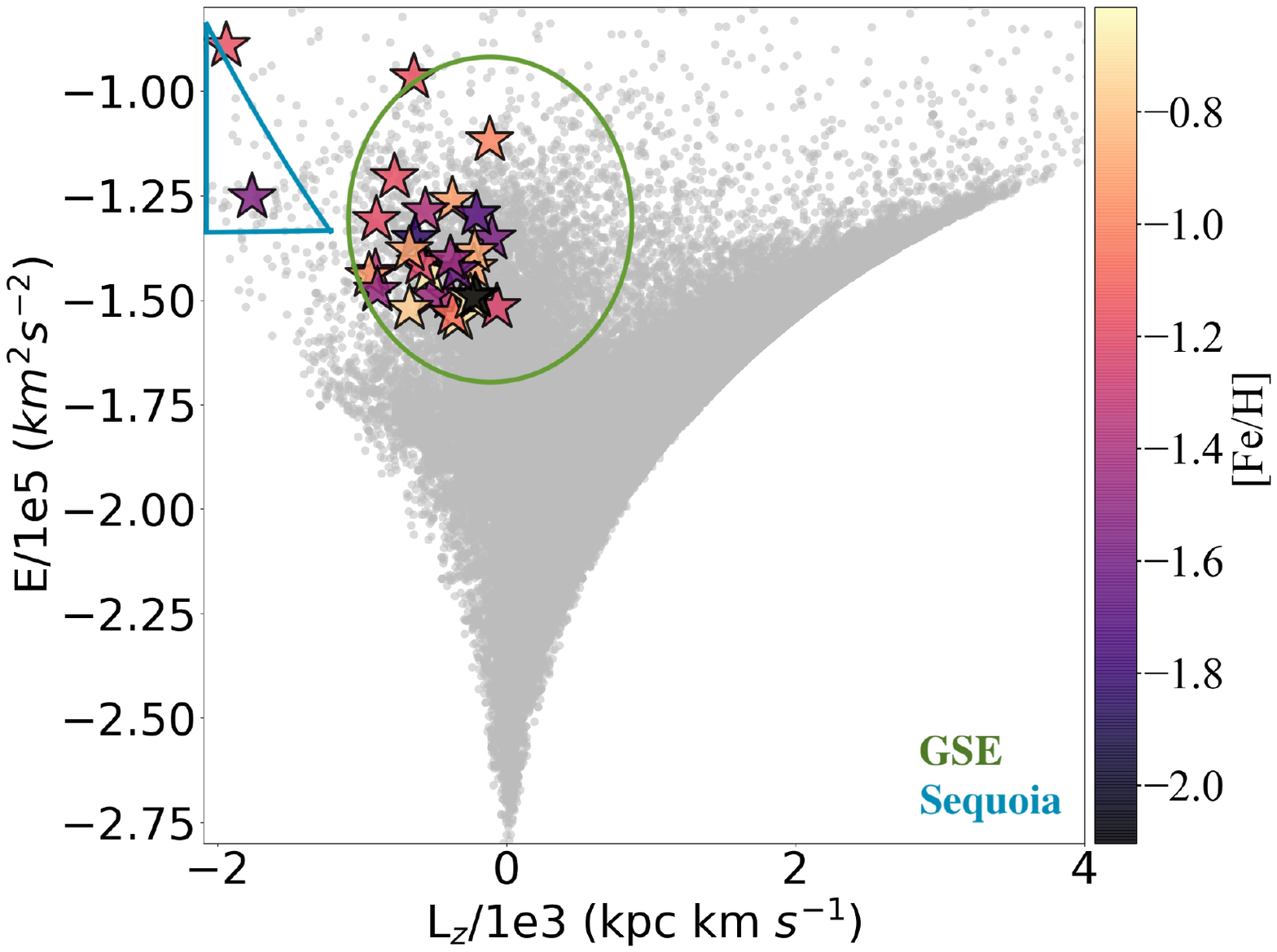}\\
\caption{L$_{z}$-E plane. Gray dots represent GALAH DR3 stars, while star markers denote the stars investigated in this study. Top panel displays the stars studied in this work color-coded by eccentricity, while bottom panel shows them color-coded by [Fe/H]. The approximate regions corresponding to GSE debris (green ellipse) and Sequoia debris (blue triangle) are indicated in the figure.}
\label{fig:dynamics}
\end{figure}

Figure~\ref{fig:dynamics} shows the L$_{z}$-E plane for our stars, represented as stellar markers, with stars from GALAH DR3 depicted in gray. In the upper panel, our stars are color-coded according to eccentricity, while in the lower panel, they are color-coded according to metallicity. Notably, our stars predominantly cluster in a specific region with L$_{z}$ close to 0, with two outliers. 

We observed that the majority of the stars exhibit high eccentric orbits, with ecc > 0.7 in most cases. However, each group contains one star with lower eccentricity, which is commented in more detail in Section \ref{subsec:peculiar_stars}.

From a dynamical perspective, comparing regions within the L${z}$-E plane associated with different progenitor galaxies across various studies poses challenges due to variations in the solar velocities, Galactic height and radius as well as Galactic potential adopted. However, based on \cite{Koppelman2019} and  \cite{horta2022chemical}, the L$_{z}$, E and eccentricity reported in this work are consistent with those of accreted stars. 

Based on the  L$_{z}$-E plane, we identify that the majority of our stars are likely associated to GES, considering the L$_{z}$ close to zero, and high E. The two outlier stars can possibly be remnants of the Sequoia galaxy, given their high E and low L$_{z}$. In Figure~\ref{fig:dynamics} we show the approximate regions were GSE debris (green ellipse) and Sequoia debris (blue triangle) are believed to be located in the L$_{z}$-E plane. Furthermore, the prevalence of high eccentricity orbits in our sample aligns with the notion that these stars are likely accreted and associated with GES and Sequoia. The low metallicity also agrees with the properties of accreted stars coming from dwarf galaxies reported so far in the literature.

\subsection{The different $\log g$ values}\label{subsec:teff_logg_results}

To investigate how similar the seismic, predicted and spectroscopic $\log g$ are and their impact on age and chemical abundance determinations (as discussed in more detail in the Sections \ref{subsection:results_ages} and \ref{subsection:abundances_vs_loggs}), we created a "golden" sample comprising 16 stars. The selection criteria for the golden sample involved ensuring consistency in seismic $\log g$ values, either through agreement between at least two of the pipelines (COR, A2Z, and BHM) considering three sigmas or when a unique value from the A2Z pipeline was available. The golden sub-sample is composed of the following stars: 405, 406, 407, 408, 409, 412, 413, 414, 415, 418, 419, 422, 424, 425, 426 and 430.

\begin{figure}
\centering
\includegraphics[width=9cm]{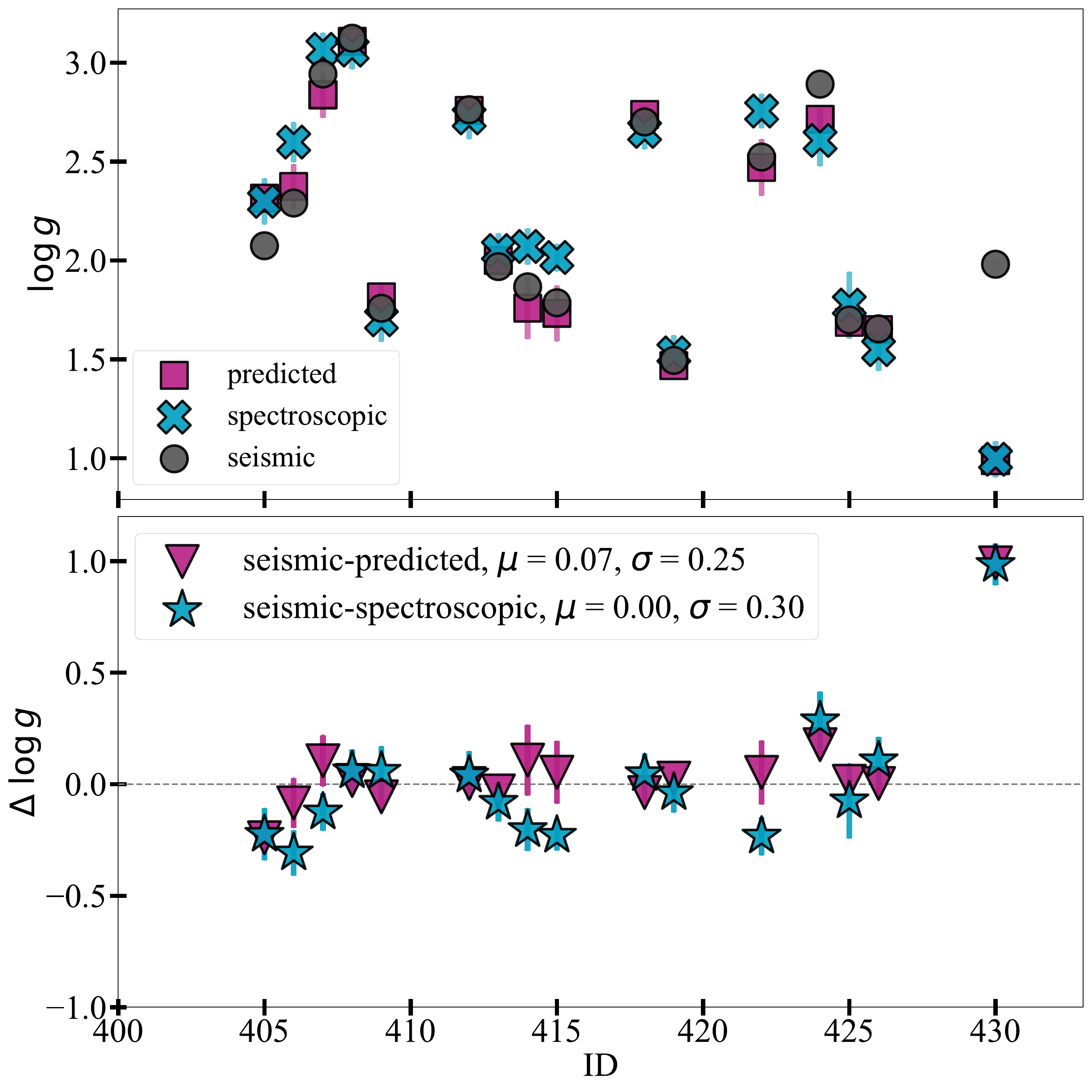}
\caption{Comparison of different $\log g$. Top panel shows the values of seismic (gray), predicted (magenta) and spectroscopic (blue) $\log g$ for each star, as well as their respective uncertainties. Bottom panel shows the difference between seismic and predicted $\log g$ (magenta), as well as between seismic and spectroscopic $\log g$ (blue). We consider as uncertainties the quadratic sum of the individual uncertainties of each case. The mean ($\mu$) as well as standard deviation ($\sigma$) of the comparison presented in the bottom panel is shown in the label.}
\label{fig:logg_comparison}
\end{figure}

In the top panel of Figure \ref{fig:logg_comparison}, we present the different $\log g$ used in this work along with their uncertainties for each golden sub-sample star. The seismic, predicted and spectroscopic values are represented in gray, magenta and blue respectively. The bottom panel of the same figure illustrates the differences between seismic and predicted $\log g$ values (magenta), as well as between seismic and spectroscopic $\log g$ values (blue). We consider the uncertainties as the quadratic sum of the individual uncertainties in each case. The mean ($\mu$) and standard deviation ($\sigma$) of the comparisons presented in the bottom panel are indicated in the label.

The mean ($\mu$) difference between the seismic $\log g$ and the predicted $\log g$ is 0.07 dex with a standard deviation ($\sigma$) of 0.25 dex. Similarly, the mean difference between the seismic $\log g$ and the spectroscopic $\log g$ is 0.00 with a standard deviation of 0.30 dex. In the case of seismic vs. predicted $\log g$, there are three cases (stars 405, 424, and 430) where the difference is not compatible with zero within 3 sigmas. In the comparison between seismic and spectroscopic $\log g$, there are two cases (stars 415 and 430) where the differences exceed what can be explained by the uncertainties within 3 sigmas. We note that the object 424 is a spectroscopic binary star \citep{pasquini1994chromospheric}, therefore it will be not be further explored in this work. Consequently, we conclude that both the predicted and spectroscopic $\log g$ are in agreement with the seismic $\log g$. However, we note that the golden sub-sample was built in a way that facilitate the agreement between the seismic and predicted values.

The uncertainties associated with each $\log g$  value vary depending on the method used to calculate it. The seismic $\log g$ represent the results with the higher precision, which mean uncertainties of the order of 0.01 dex, while the spectroscopic $\log g$ represent the less precise case, with mean uncertainties of 0.1 dex, which is 10 times larger then the typical seismic ones. The predicted $\log g$ has a uncertainty distribution with a mean of 0.06 dex, which is close to the precision obtained using spectroscopy.

\subsection{Stellar ages}\label{subsection:results_ages}

To assess the impact of the use of seismic information on age determination, we employed the \texttt{BASTA} code considering the three cases described in Section \ref{subsec:ages_methods}, namely seismic, predicted, and spectroscopic. From the sixteen stars that had measured $\nu_\mathrm{max}$ and $\Delta$$\nu$ and where part of the golden sample, only thirteen had ages successfully measured in the three cases.  

We identify the impact of using seismic information in the statistical uncertainties obtained in each case of study as provided by \texttt{BASTA}. When calculating ages using seismic information, the typical uncertainty found is of the order of 10.1 percent. When using the predicted $\log g$, the uncertainties represent about 18.5 percent. Finally, when using spectroscopic $\log g$, the typical uncertainty represent 28.6 percent of the measurement in this work. Therefore, our results suggest that using seismic information when calculating ages of metal-poor accreted stars candidates can drastically improve the stellar age precision when considering statistical uncertainties, which illustrates the potential of \texttt{TESS} data for Galactic Archaeology studies.

\begin{figure}
\centering
\includegraphics[width=8cm]{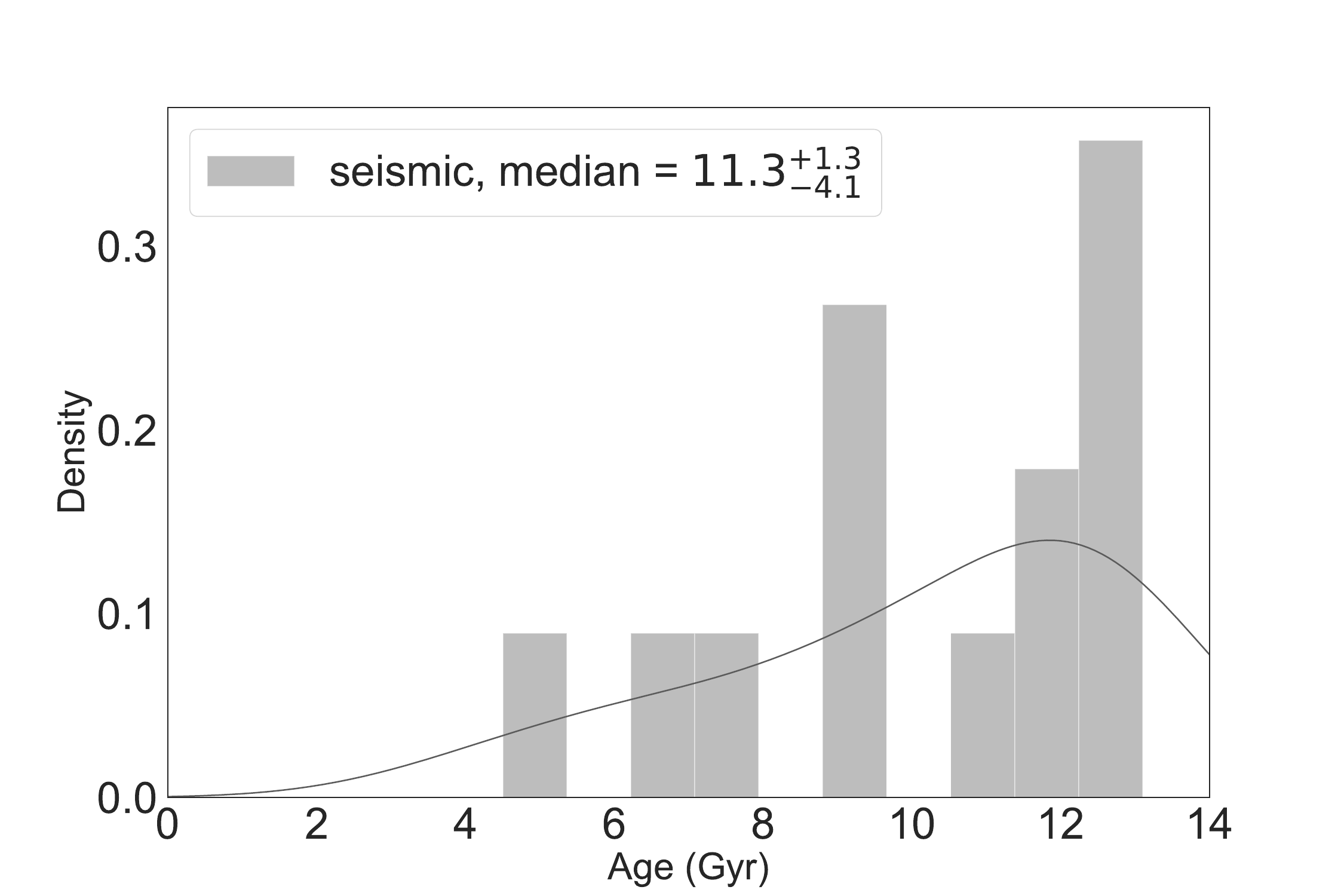}
\includegraphics[width=8cm]{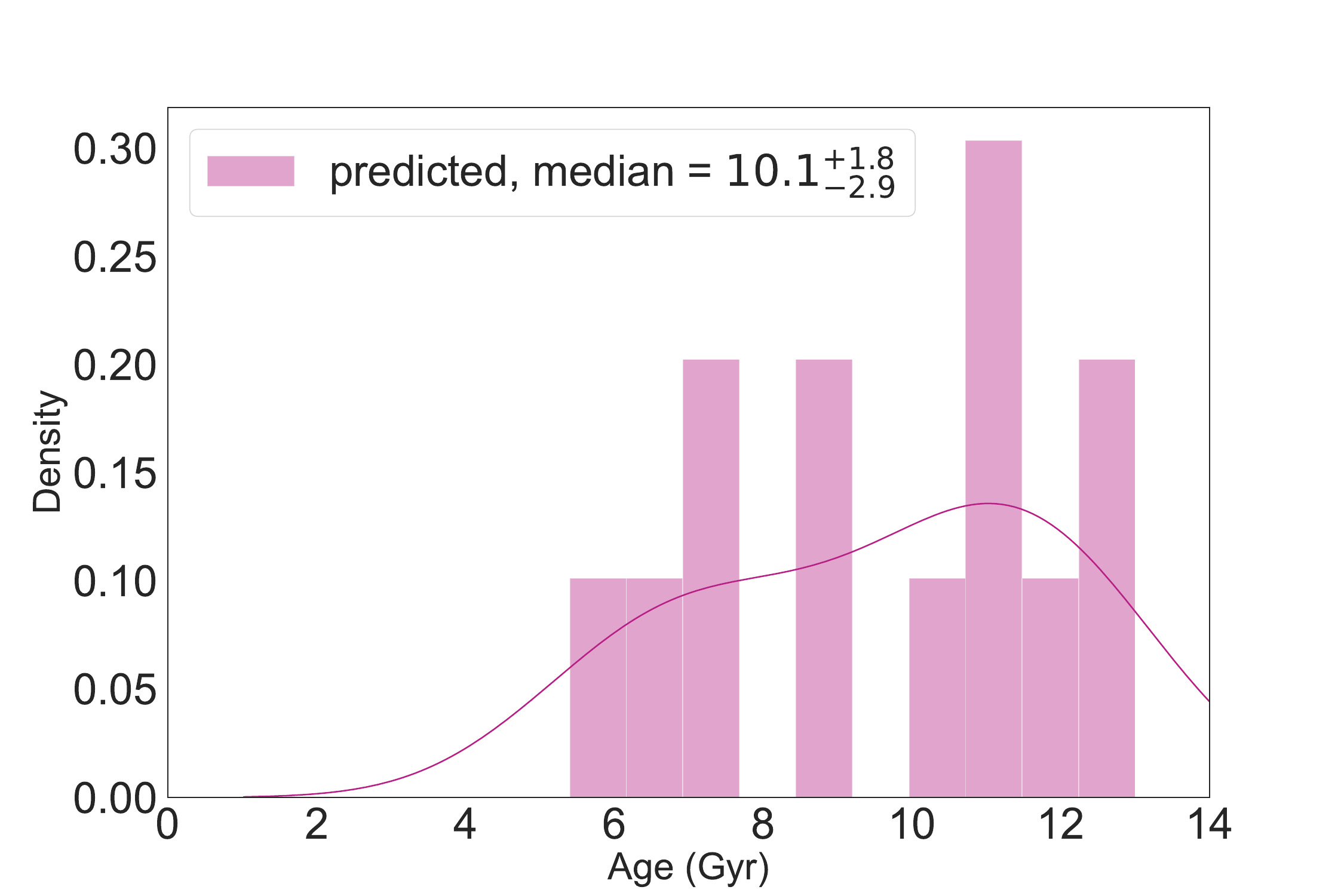}
\includegraphics[width=8cm]{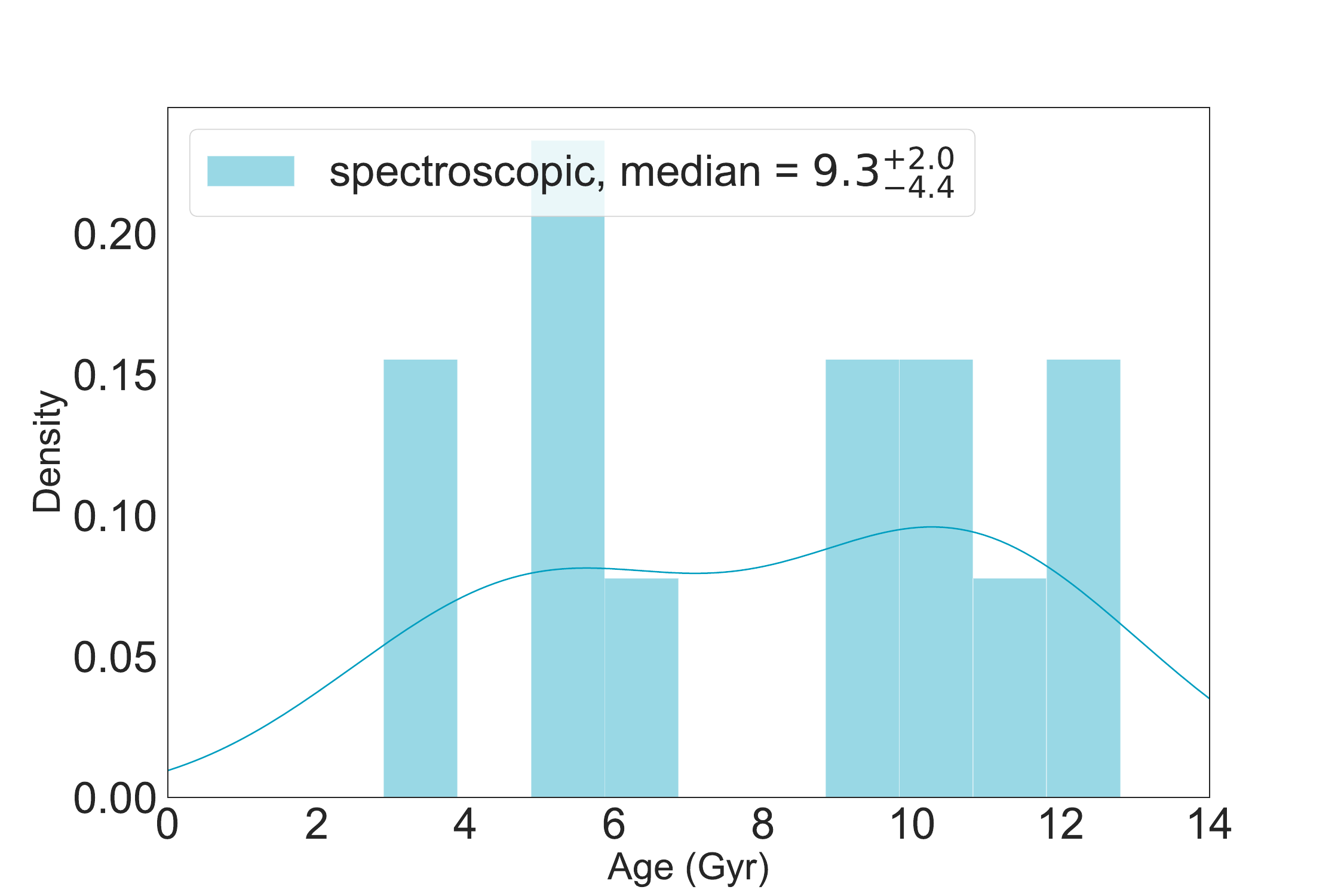}\\
\caption{Age distribution for the three cases studied in this work using \texttt{BASTA}. Top panel (gray) shows stellar ages calculated using the seismic information $\nu_\mathrm{max}$ and $\Delta$$\nu$. Middle panel (pink) shows stellar ages calculated using the predicted $\log g$. Bottom panel (blue) shows the stellar ages calculated using the spectroscopic $\log g$. Median values as well as  lower and upper uncertainties are indicated in the label.}
\label{fig:ages_seismo_predicted_spectroscopic}
\end{figure}

In Figure \ref{fig:ages_seismo_predicted_spectroscopic} we compare the ages obtained when using seismic information (including $\nu_\mathrm{max}$ and $\Delta$$\nu$; top panel) and when using the predicted (middle panel) and the spectroscopic (lower panel) $\log g$. Additionally, the ages obtained for individual stars as well as their uncertainties considering the seismic, predicted and spectroscopic $\log g$ are provided in Table \ref{tab:ages}.

\begin{table*}[h!]
\centering
\label{tab:ages}
\begin{spacing}{1.75}
\caption{Ages of individual stars calculated using seismic, predicted and spectroscopic $\log g$, when results for the three cases studied were available.}
\begin{tabular}{|r|r|r|r|r|r|r|}
\hline
  \multicolumn{1}{|c|}{ID} &
  \multicolumn{1}{c|}{Age (Gyr)} &
  \multicolumn{1}{c|}{Age (Gyr)} &
  \multicolumn{1}{c|}{Age (Gyr)} \\
  \multicolumn{1}{|c|}{} &
  \multicolumn{1}{c|}{seismic} &
  \multicolumn{1}{c|}{predicted} &
  \multicolumn{1}{c|}{spectroscopic} \\
\hline
        405 & $12.6_{-0.8}^{+0.7}$ & $10.1_{-0.9}^{+1.2}$ & $5.8_{-1.8}^{+2.6}$ \\
        406 & $11.9_{-1.6}^{+1.2}$ & $7.2_{-3.4}^{+3.7}$ & $2.9_{-1.5}^{+2.9}$ \\
        407 & $4.5_{-0.5}^{+1.1}$ & $6.4_{-1.0}^{+1.9}$ & $5.0_{-2.8}^{+1.5}$ \\
        408 & $9.6_{-0.3}^{+0.4}$ & $10.8_{-0.7}^{+0.6}$ & $10.5_{-2.4}^{+0.9}$ \\
        409 & $9.5_{-1.3}^{+1.1}$ & $8.7_{-3.4}^{+1.5}$ & $9.3_{-1.7}^{+1.3}$ \\
        412 & $7.2_{-0.9}^{+1.1}$ & $7.2_{-1.0}^{+1.3}$ & $6.8_{-1.4}^{+2.3}$ \\
        413 & $12.2_{-0.9}^{+0.8}$ & $11.9_{-1.2}^{+0.9}$ & $11.9_{-1.6}^{+0.9}$ \\
        414 & $11.3_{-1.5}^{+1.5}$ & $11.4_{-1.9}^{+1.6}$ & $9.9_{-2.9}^{+2.2}$ \\
        415 & $12.6_{-0.6}^{+0.7}$ & $13.0_{-0.9}^{+0.6}$ & $12.8_{-1.1}^{+0.7}$ \\
        418 & $6.5_{-0.6}^{+1.0}$ & $5.4_{-0.6}^{+0.5}$ & $5.3_{-0.7}^{+0.5}$ \\
        419 & $13.1_{-0.7}^{+0.5}$ & $12.3_{-1.4}^{+0.9}$ & $11.3_{-2.0}^{+1.5}$ \\
        422 & $12.6_{-1.0}^{+0.7}$ & $11.0_{-3.2}^{+2.2}$ & $3.6_{-1.2}^{+1.8}$ \\
        426 & $9.3_{-1.1}^{+1.7}$ & $9.1_{-1.3}^{+2.3}$ & $9.5_{-1.6}^{+2.6}$ \\  
        \hline
\end{tabular}
\end{spacing}
\end{table*}

The age distribution for the seismic case goes from 4.5 Gyr to 13.1 Gyr, with a median of 11.3 Gyr, a lower uncertainty of 4.1 Gyr and a upper uncertainty of 1.3 Gyr. From the 13 stars considered, only two stars are younger than 7 Gyr. Therefore, considering asteroseismic information, 85 percent of our sample can be considered old, resulting in the age distribution presented in Figure \ref{fig:ages_seismo_predicted_spectroscopic}.

When considering the ages calculated using the predicted $\log g$, the distribution goes from 5.4 Gyr to 13.0 Gyr, with a median of 10.1 Gyr, a lower uncertainty of 2.9 Gyr and a upper uncertainty of 1.8 Gyr. Repeating what was observed in the ages calculated using seismic information, 85 percent of our sample contains stars older than 7 Gyr according to our results. 

The ages calculated using the spectroscopic $\log g$ have a median of 9.3 Gyr, a lower uncertainty of 4.4 Gyr and a upper uncertainty of 2.0 Gyr, and a range between 2.9 Gyr and 12.8 Gyr. From the 13 stars, 6 have ages below 7 Gyr according to our analysis, representing 46 percent of our sample. Considering the significant size of the young sub-sample, the age distribution appears to be bimodal.

Ages calculated using seismic information are generally higher than ages calculated using the predicted and the spectroscopic $\log g$. When compared with the age distribution calculated using the predicted $\log g$, seismic ages appear to be 1.2 Gyr older, while when compared with the age distribution calculated using spectroscopic $\log g$ it appears to be 2.0 Gyr older. We note that the age distribution calculated using the predicted $\log g$ is similar to the one computed with seismic information, since the $\nu_\mathrm{max}$ and $\Delta$$\nu$ used in the seismic case are correlated with the predicted $\log g$ (see Section \ref{sec:data} and Section \ref{subsec:teff_logg_methods}.).

However, the difference in the median age in the three cases studied is small. Considering the uncertainties, we can consider the median ages are compatible within one sigma among all the three cases studied. We also note that the difference among the median values is of the order of the typical uncertainty of ages of the individual stars explored in this work, which mostly varies between 0.5 Gyr and 2.5 Gyr, indicating that the age resolution is currently not big enough to discern small variations in the median values of the distributions.

As an attempt to better discern the behavior of the three age distributions, we evaluated them using a Bayesian approach. In this method, we considered each individual age measurement as a Gaussian distribution and reassessed the overall age distribution. Using this Bayesian approach, we obtained a seismic distribution with a median of 10.2 Gyr and uncertainty of 0.2 Gyr. The predicted distribution showed a median of 9.3 Gyr with uncertainty of 0.3 Gyr. Lastly, the spectroscopic distribution resulted in a median of 7.8 Gyr, with uncertainty of 0.4 Gyr. We observed that with the Bayesian approach, the median age is lower in all cases, the uncertainties are smaller for all distributions, and while the median ages are still compatible within three sigmas when considering the seismic and predicted cases, the spectroscopic median does not agree with the seismic median within three sigmas.

The dominant old age distribution found in the seismic ages and the ages calculated using the predicted $\log g$, agree well with previous works in the literature. As an example,  \cite{das2020ages} calculated ages using a Bayesian isochrone pipeline for a structure composed of accreted star candidates named the \texttt{blob} and found an age span from 8 to 13 Gyr, which is comparable with results found in this work. \cite{montalban2021chronologically} calculated ages of [Mg/Fe]-poor stars with high eccentric orbits that are accreted stars candidates using seismic data, and found a mean age distribution of 9.7 $\pm$ 0.6 Gyr. \cite{feuillet2021selecting} found a mean age distribution for accreted stars from the GES progenitor of 10-12 Gyr. As previously stated, \cite{borre2022age} calculated ages for accreted stars using \texttt{Kepler} and \texttt{K2} data, and found an age distribution of $9.5_{-1.3}^{+1.2}$. We note that even though the ages calculated using  the spectroscopic $\log g$ appears to be younger than the other cases studied, 54 percent of that sample is composed of stars that appears to be old, which agree with the literature. 

In our results, we found that about 15 percent of the ages calculated using the seismic information or the predicted $\log g$ appears to be younger than 7 Gyr, in contrast with the 46 percent found when using the spectroscopic $\log g$. The young ages are in contrast with the low metallicity, since according the delayed SNIa scenario, metal-poor stars tends to be older than metal-rich stars. 

We note that in \cite{grunblatt2021age}, the authors calculated ages of halo substructures using among other codes, \texttt{BASTA}. In that work they studied 10 red giant stars using kinematic, asteroseismic and spectroscopic data, and a sub-sample of them were associated to GES. In that work, the probability density function of the age distribution obtained using \texttt{BASTA} spanned a wide range, going from 1 Gyr up to ages as old as 13 Gyr. In comparison to their work, the results we find for ages calculated using seismic data appear to be older, but we also found a small population of young stars similar to their work. However, we note that the setup used in this work to apply \texttt{BASTA} to our data is different from the one used in \cite{grunblatt2021age}.  Similarly, using data from the Large Sky Area Multi-Object Fibre Spectroscopic Telescope (\texttt{LAMOST}, \citealt{zhao2012lamost}), \cite{horta2024stellar} found a population of accreted stars with a mean age of 11.6 Gyr, but containing a tail of intermediate age stars with ages between 6 and 9 Gyrs. Therefore, accreted stars with lower ages has also been reported before in the literature.

Based on our results, we interpret that the use of seismic information impacts the age determination. Among the three studied age sets, seismic ages tend to be older. We note that this can be caused by the configuration chosen when using \texttt{BASTA} in this work.
The age set calculated using seismic ages are the one providing older ages, therefore is the one we consider more trustworthy. While  the impact in chemical abundances when using or not using seismic information is small (see Section \ref{subsection:abundances_vs_loggs} for details), the same is not valid for ages. Therefore, age determination with or without seismic information, requires careful determination.

\begin{figure}
\centering
\includegraphics[width=9cm]{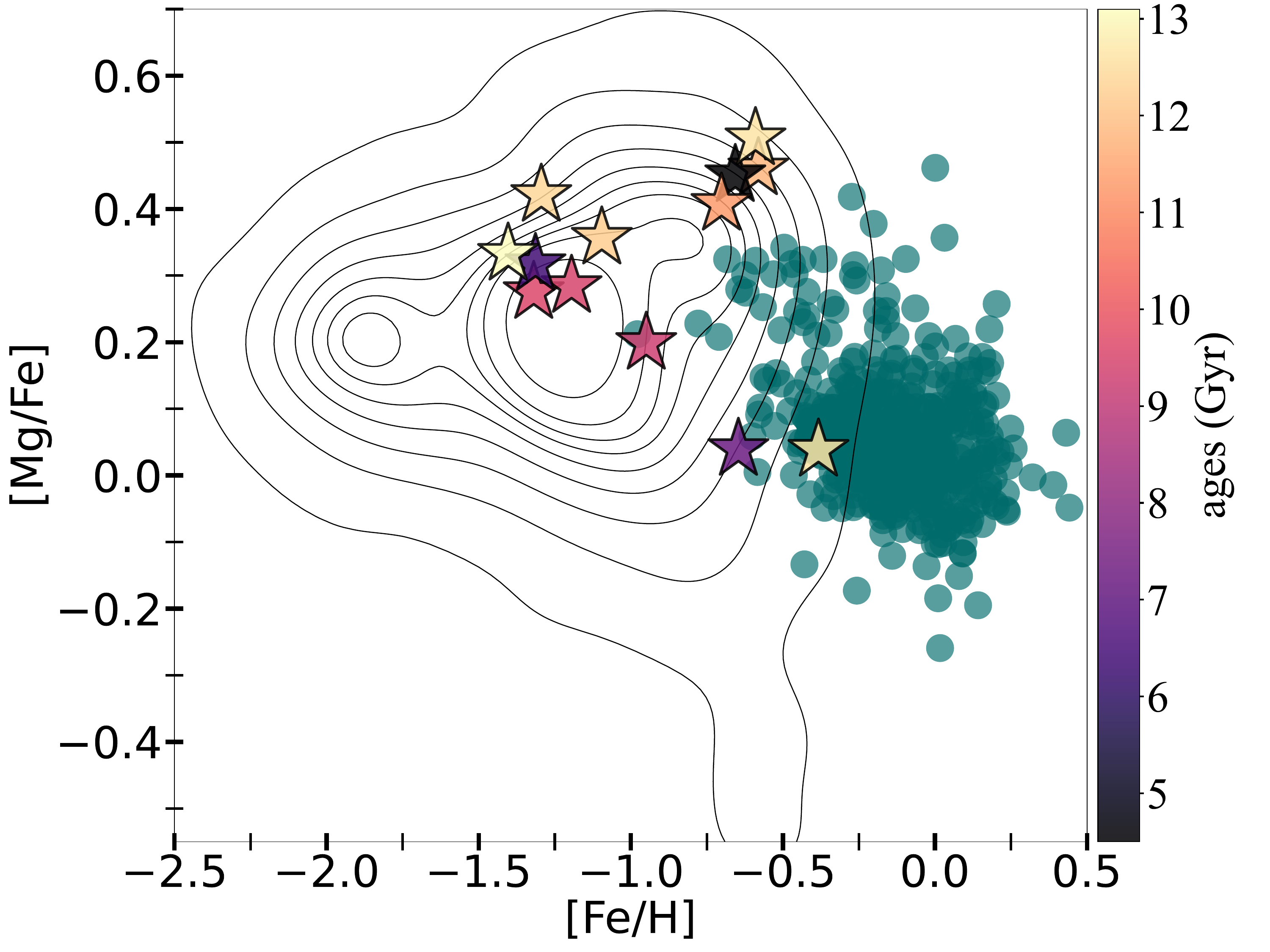}\\
\caption{[Mg/Fe] vs. [Fe/H]. The stars in this work are presented as stellar markers, while other stars from \cite{mackereth2021prospects} are represented as teal dots (from GALAH DR3 data). We also show as black contour lines the kernel density estimate of metal-poor stars with retrograde motion from GALAH DR3 survey. The stars presented in this work are color-coded according to \texttt{BASTA} seismic stellar ages.}
\label{fig:ages_feh_dist}
\end{figure}

In Figure \ref{fig:ages_feh_dist}, we present a [Mg/Fe] vs. [Fe/H] plane, with metal-rich stars from \cite{mackereth2021prospects} represented as teal dots and stars presented in this work represented as stellar symbols color-coded according to the seismic stellar age. As a metal-poor control sample, we present the as contour lines the kernel density estimate of metal-poor stars with retrograde motions from GALAH DR3 survey, that have stellar parameters similar to the ones of the stars we studied in this work. The chemical abundances for the metal-rich stars from \cite{mackereth2021prospects} were obtained from GALAH DR3. As discussed above, our sample is composed of mostly old stars, with only two objects younger than 7 Gyr.

The stars which appear to be younger have a chemical pattern consistent with those of old stars. One possible scenario to explain the young ages is that the stars could be a product of binary evolution, following the discussion presented \cite{de2022j01020100}. The star could also be a young $\alpha$-rich star \citep{chiappini2015young,martig2015young,jofre2016cannibals}. Young $\alpha$-rich stars are believed to be a product of binary evolution or stars formed near the Galactic bar co-rotation, which is a region where gas could be kept inert for long periods of time. More recent discussions on young $\alpha$-rich star can be found at \cite{jofre2023cannibals,cerqui2023stragglers,grisoni2024k2}. However, while this can be a possibility to explain the two young stars found calculating ages using seismic information and predicted $\log g$, the larger population of young stars found when using spectroscopic $\log g$ are more likely to be explained by a necessity to include seismic information to better constrain the ages of these objects.

\subsection{Chemical abundances based on different $\log g$}\label{subsection:abundances_vs_loggs}

\begin{figure*}
\centering
\includegraphics[width=19.5cm]{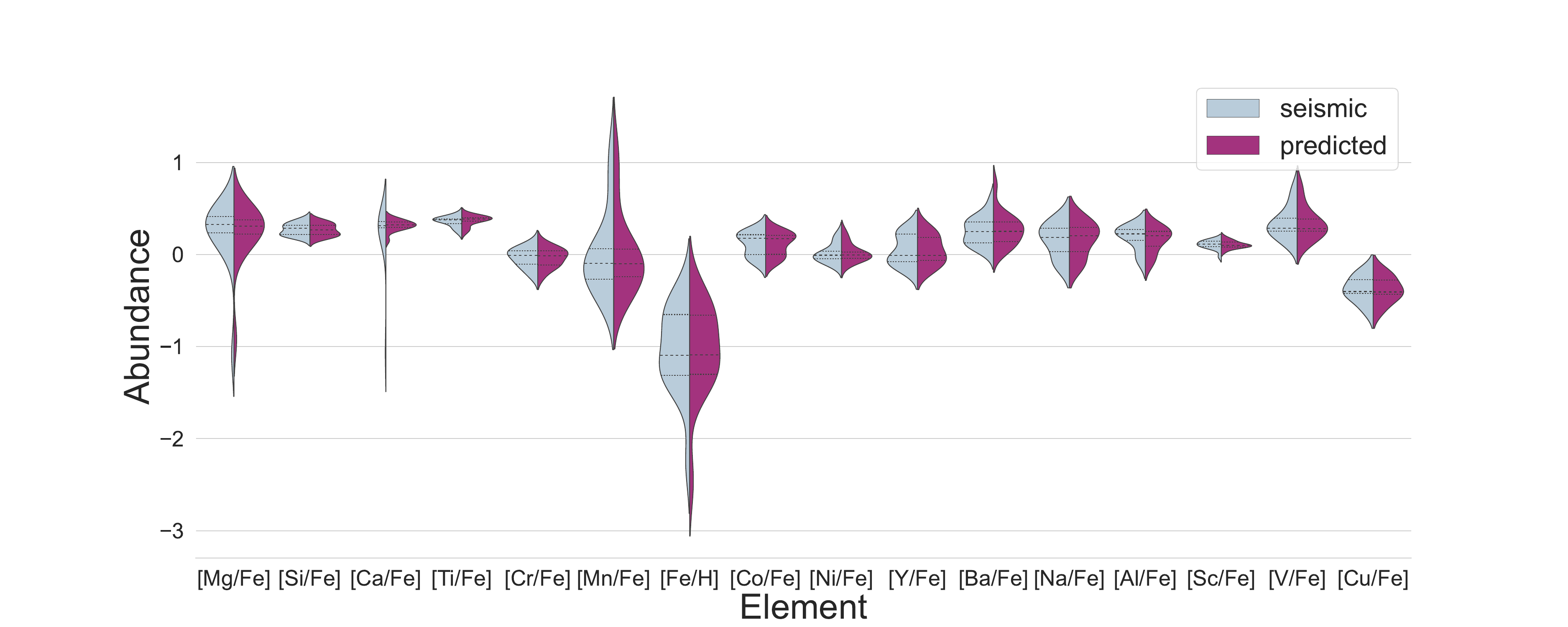}
\includegraphics[width=19.5cm]{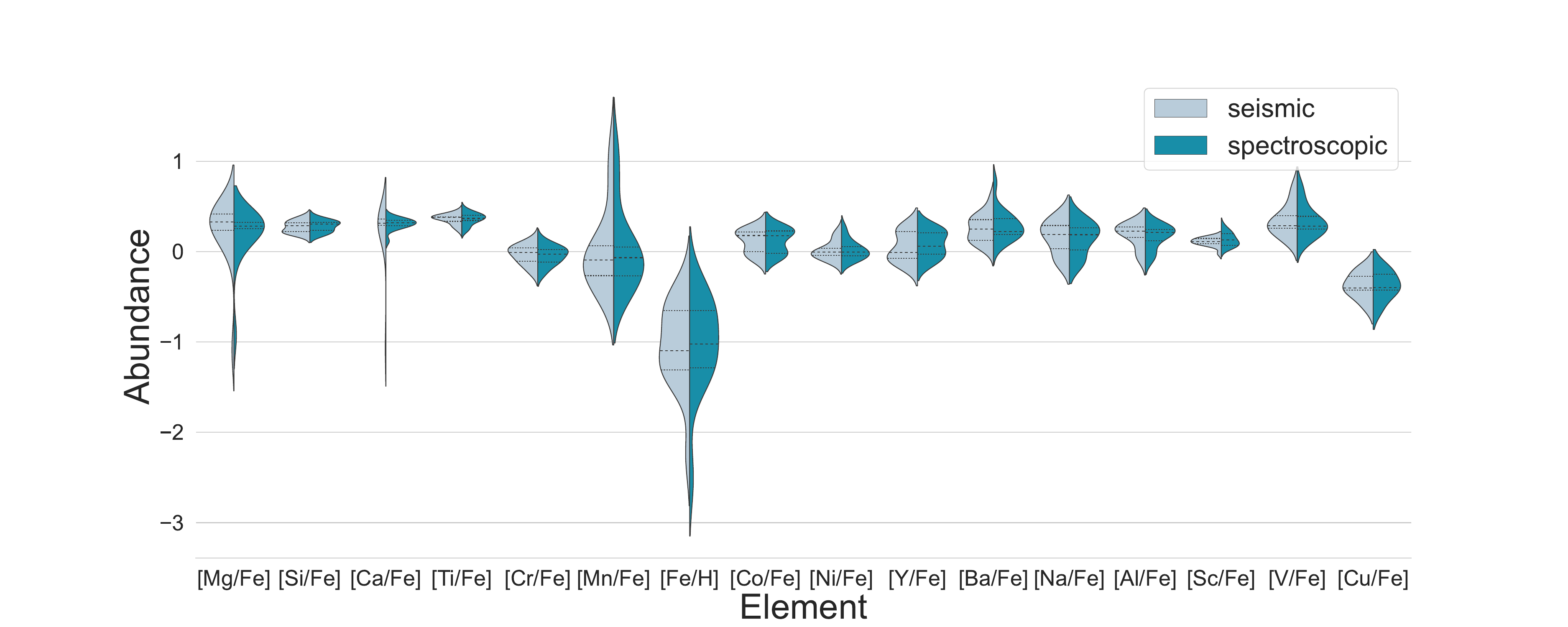}
\caption{Comparison of chemical abundance distribution obtained adopting the different $\log g$. Top panel shows the chemical distribution per element when using the seismic $\log g$ (gray) and the predicted $\log g$ (magenta). Bottom panel shows the chemical distribution per element when adopting the seismic $\log g$ (gray) and the spectroscopic $\log g$ (blue). The center dashed line represents the median of the distributions, while the other lines neighboring it shows the data quartiles. The chemical abundances are ordered according to main nucleossynthetic channel: $\alpha$-capture, Iron peak, neutron-capture and Odd-Z. Within each nucleossynthetic channel, the chemical elements are organized according to atomic number.}
\label{fig:violin_predicted_seismo_spectroscopic}
\end{figure*}

In the top panel of Figure \ref{fig:violin_predicted_seismo_spectroscopic} we present a comparison between the chemical abundance distribution obtained using seismic $\log g$ and those obtained using the predicted $\log g$.  Overall, we can deduce that the chemical abundance distribution is very similar when considering different $\log g$. However, we note that calcium (Ca) appears to be more affected by the choice of $\log g$. 

The distribution of calcium (Ca) also exhibits variations between the two results. Ca, determined with seismic $\log g$, shows a broader, flatter distribution that covers a wider range compared to the distribution obtained using predicted $\log g$. Ca abundances were determined using only neutral lines, but Ca is an element known to be highly influenced by surface gravity.

In the bottom panel of Figure \ref{fig:violin_predicted_seismo_spectroscopic} we present a comparison between the chemical abundances obtained using seismic $\log g$ and those obtained using the spectroscopic $\log g$. Similar to the previous comparison, the overall shape and range of the chemical abundance distributions is the same when we use the seismic $\log g$ and the spectroscopic $\log g$. Differences are observed in the distribution of Ca, and Sc.

Ca, in this case, displays a two-peaked distribution, contrasting with the flatter and broader distribution obtained using seismic $\log g$. It is important to note that both distributions of Ca abundances were estimated using only neutral lines, highlighting the impact of surface gravity on this element. In the case of Sc, while the seismic distribution is narrow, the distribution of Sc determined with spectroscopic $\log g$ is broader with a tail.

We also note that even though the range and general shape of the distribution of [Fe/H] is the same in the two chemical sets, the stars appears to be slightly metal-poorer (about 0.1 dex) when the abundance is determined using the seismic $\log g$. When calculating [Fe/H] we used neutral and ionized lines, with neutral lines being the large majority.

Additionally, across all the chemical sets explored, distributions for Si, Co, Ni, Y, and Al consistently show two peaks. A few stars in the sample exhibit relatively high abundances of Mn and V, while the overall sample is rich in Ti. The stars also show a wide range of Mn abundances and low Cu abundances.

\begin{figure}
\centering
\includegraphics[width=8cm]{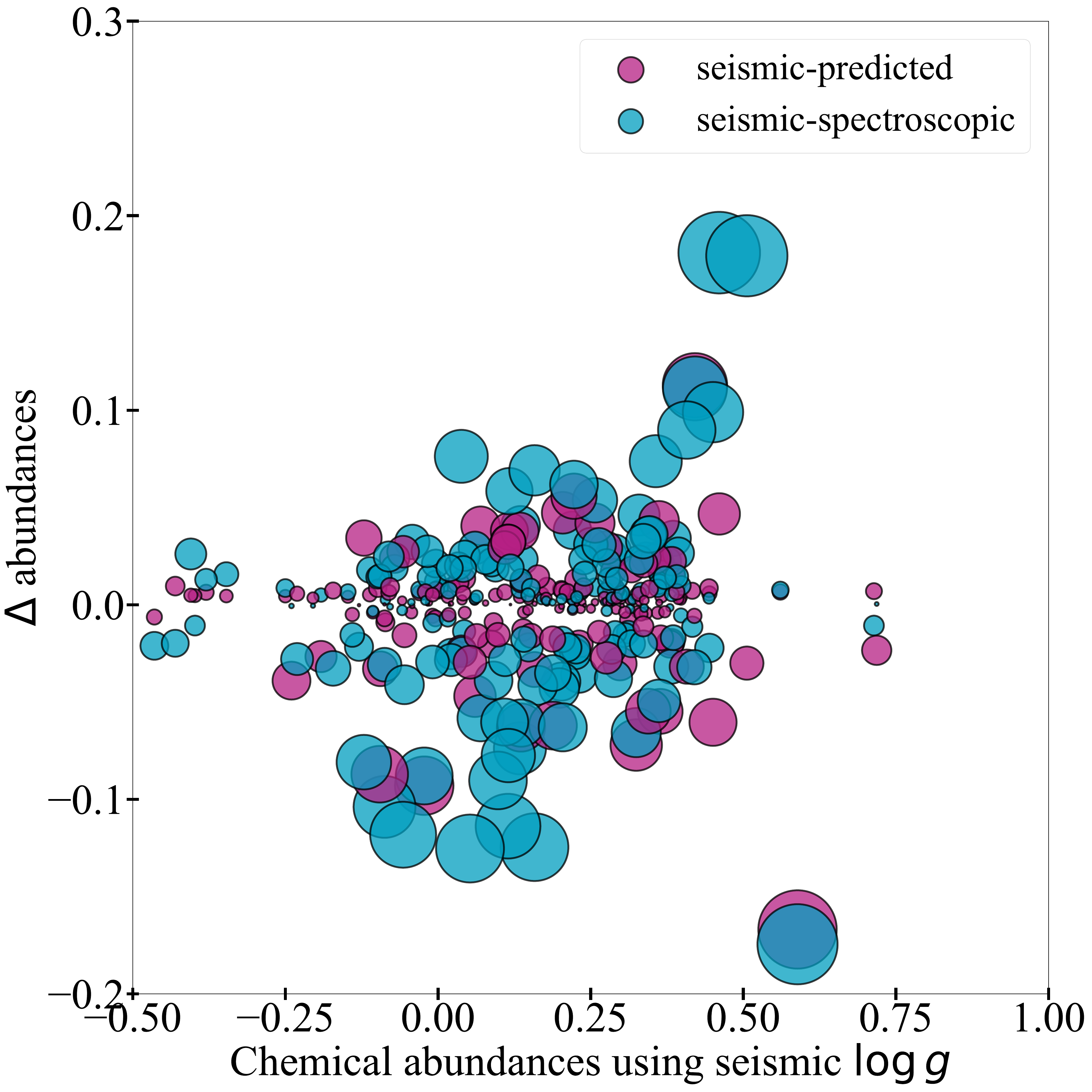}
\caption{Difference in chemical abundances for all elements ($\Delta$ [X/Fe]) as a function of chemical abundances determined adopting the seismic $\log g$. The difference between abundances estimated using seismic and predicted $\log g$ are shown in magenta, while the difference between the abundances determined adopting seismic and spectroscopic $\log g$ are shown in blue. The marker size is proportional to the difference between the $\log g$ of each case.}
\label{fig:delta_abundances_logg}
\end{figure}

Considering the results from Section \ref{subsec:teff_logg_results}, where the seismic, predicted and spectroscopic $\log g$ values were generally compatible, it is not surprising that the overall chemical abundances distribution is as similar as seen in Figure \ref{fig:violin_predicted_seismo_spectroscopic}. To better assess the relationship between the difference in chemical abundances ($\Delta$ [X/Fe]) and the difference in $\log g$ we present Figure \ref{fig:delta_abundances_logg}. This figure shows $\Delta$ [X/Fe] as a function of the abundances estimated using the seismic $\log g$, with marker sizes proportional to the $\log g$ difference for each case. The $\Delta$ [X/Fe] were calculated subtracting the chemical abundances determined using the predicted $\log g$ from the ones estimated using the seismic $\log g$ (magenta) and subtracting the chemical abundances determined adopting the spectroscopic $\log g$ from the chemical abundances estimated adopting the seismic $\log g$ (blue). We note that $\Delta$ [X/Fe] encompass all chemical elements studied in this work.

Figure \ref{fig:delta_abundances_logg} illustrates that the difference in chemical abundances estimated using the predicted or spectroscopic $\log g$ compared to abundances determined using seismic $\log g$ is correlated with the difference in $\log g$. Therefore, when the difference in the $\log g$ for a certain star is substantial, it results in a notable difference in chemical abundances. However, $\Delta$ [X/Fe] is mostly distributed between -0.1  < $\Delta$ [X/Fe] < 0.1 dex, which is usually small enough to be accounted by the uncertainties in the chemical abundances.

We note that chemical abundances determined with or without seismic information are largely consistent, suggesting that the availability of seismic data for accreted stars does not significantly impact abundance determination using our methodology, when considering the chemical elements studied in this work. However, it is important to note that using seismic information, the uncertainty of $\log g$ can be dramatically decreased. Nevertheless, minor impact in the abundances of [Fe/H], [Ca/Fe], and [Sc/Fe] were observed, therefore these chemical elements might need more careful treatment.

\subsection{Chemical abundance distribution}\label{subsec:chemical_abundance_distribution}

\begin{figure*}
\centering
\includegraphics[width=17.5cm]{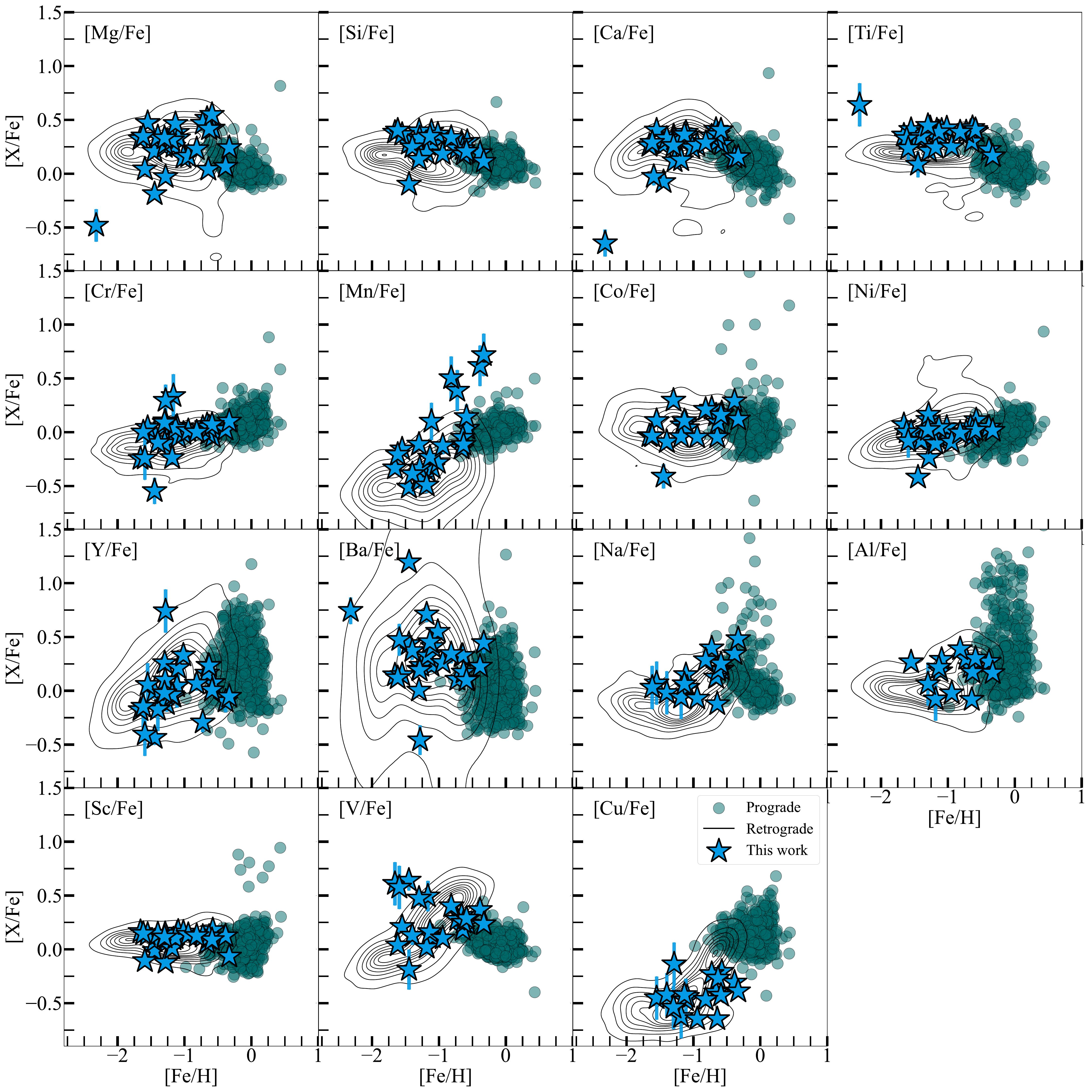}\\
\caption{Chemical abundance distribution as a function of [Fe/H]. Stars in blue are those studied in this work. The chemical abundances are those obtained using the spectroscopic $\log g$. The teal dots (prograde stars from \citealt{mackereth2021prospects}) and black contours ( GALAH DR3 retrograde metal-poor stars) are the same control sample stars adopted in Figure \ref{fig:ages_feh_dist}. The chemical elements are ordered according to their main nucleossynthetic channel: $\alpha$-capture, Iron peak, neutron-capture and Odd-Z. Within each nucleossynthetic channel, the chemical elements are organized according to atomic number.}
\label{fig:abundance_space}
\end{figure*}

Considering the satisfactory agreement among the different $\log g$ and chemical abundances estimated using the different $\log g$ previously discussed, we determined chemical abundances for the 29 stars in our sample using the spectroscopic $\log g$, excluding the spectroscopic binary 424. . In Figure \ref{fig:abundance_space}, we present the distribution of the chemical abundances of the stars analysed in this work, represented by blue stellar markers. Additionally, we include a control sample of GALAH DR3 prograde stars, shown as filled teal circles, that were also studied in \cite{mackereth2021prospects}. We chose this set of stars as comparison since they are also TESS stars for which optical spectra are available. They are more metal-rich, and from their kinematics, it is likely that they are objects formed \textit{in situ}.  We also considered a metal-poor counter-rotating control sample of GALAH DR3 stars that have atmospheric parameters similar to the stars explored in this work.
This allow us to compare the chemical pattern of stars born \textit{in} and \textit{ex-situ}.

A significant portion of our stars displays [Fe/H] < -0.8, which is consistent with the metal-poor nature of the majority of accreted stars, as reported in previous studies such as \cite{Helmi2018,matsuno2019origin,das2020ages,feuillet2021selecting}. Nonetheless, a minority of stars exhibits [Fe/H] > -0.8, which deserves further discussion in Section \ref{subsec:peculiar_stars}.

$\alpha$ elements such as Mg, Ca, Si, and Ti (which is not an $\alpha$-capture element, but behaves like one) are good tracers of core collapse supernovae \citep{timmes1995galactic,kobayashi2006galactic,kobayashi2020origin}, since they are mostly produced in massive stars and later ejected by these kind of supernovae. However, we note that about 22 percent of Si and 39 percent of Ca come from SN Ia, following the work of \cite{kobayashi2020origin}. Additionally, if other types of supernovae are considered (e.g. faint-supernovae), it is possible that a larger fraction of $\alpha$ elements are produced by other mechanisms \citep{nomoto2009nucleosynthesis}.
Our stars present a high [$\alpha$/Fe] abundance at low metallicities ($\mathrm{[Fe/H]} < -1.0$), which is typical for halo and thick disk stars. There is one star in our sample with a lower enhancement of $\alpha$-elements, which may suggest it originated in a lower-mass system than the Milky Way and was later accreted \citep{nissen2010two}. This star is better discussed in Section \ref{subsec:peculiar_stars}. The generally high [$\alpha$/Fe] values imply that the progenitor galaxy had the necessary mass and star formation rate to produce enhanced [$\alpha$/Fe] abundances, consistent with findings in the literature regarding the abundance of $\alpha$ elements in GSE stars (see \citealt{Helmi2018, carrillo2022detailed}).

We observe that the stars are metal-poor and [$\alpha$/Fe]-rich, indicating that if the stars are accreted objects, their progenitor galaxy was likely a massive system with a large star formation rate (SFR). Studies such as \cite{tolstoy2009star,hill2019vlt,ji2019chemical} had shown that the chemical pattern of stars from dwarf galaxies varies according to the mass of the system, and that a like intense star formation rate might cause the gas to be enrich toward the metal-rich regime. Additionally, the high [$\alpha$/Fe] abundance indicate numerous SNII events. While the typical abundance of $\alpha$  elements in metal-poor Milky Way halo and thick disk stars can be of the order of 0.4 dex (e.g. \citealt{ishigaki2012chemical,hawkins2015using}), the abundances of these elements can be much lower than these value in lower mass galaxies. In the Large Magellanic Clouds, Fornax and Sculptor dwarf galaxies, for example, the abundances of $\alpha$ elements reported in works such as \cite{shetrone2003vlt,letarte2007chemical,van2013chemical,hill2019vlt} can be below zero for moderately metal-poor stars.

Iron-peak elements such as nickel (Ni), chromium (Cr), cobalt (Co) and manganese (Mn) are mostly produced by Type Ia supernovae (SNIa), making them valuable tracers of historic processes happening on longer timescales. In our sample, the [Ni/Fe] and [Co/Fe] distributions are  generally flat around the solar value. [Cr/Fe] shows a slight increase with the metallicity, which is in contrast to the slight decrease reported by \cite{carrillo2022detailed}. This could be attributed to NLTE effects, since Cr is sensitive to it a low metallicities. We also observe an increase in Mn abundance with [Fe/H]. The metal-poor accreted star candidates have lower Mn abundances than the prograde stars, as expected due to the metallicity effect. We also note that the wide range of [Mn/Fe] abundance with few stars that are Mn-rich, indicates a fast star formation rate, since multiple SNIa may have enriched the gas from a low-Mn to a Mn-rich condition. 

Neutron capture elements such as barium (Ba), and yttrium (Y) are very important to Galactic Archaeology as they reflect the impact of s-process chemical enrichment in the Milky Way chemical evolution. We do not find a Ba abundance variation associated with [Fe/H], and find a few mildly Ba-enhanced stars. Y shows a large spread similar to what is observed with Ba, and two possible sequences are observed. However, these sequences may be influenced by our limited sample size. We also found a mildly Y-enhanced star.

The observed mild Ba enrichment, implies that neutron-capture elements played a significant role in shaping the chemical evolution of the progenitor galaxy. The abundance patterns of these elements may provide valuable insights into the original environments of the progenitor galaxies, star formation timescales, and chemical evolution pathways.

Odd-Z and light elements such as vanadium (V), copper (Cu), scandium (Sc), sodium (Na) and aluminum (Al) are formed by different mechanisms and can provide important information for Galactic Archaeology. We observe an increase of V and Cu with [Fe/H], along with a few stars that are strongly V-enhanced with abundances of [V/Fe] > 0.5 dex. The stars studied in this work appear to be Cu-poor when compared with the stars likely formed \textit{in situ}. Sc exhibits a slight decrease with [Fe/H], and our stars generally have high Al abundances, although three stars appear to be Al-poor. This is interesting because many works in the literature use APOGEE to explore accreted stars (e.g. \citealt{das2020ages}) that in general have low [Al/Fe], but we do not observe the same in the optical. This could be caused by systematic effects between the optical and the infrared caused by differences in the atomic data adopted. 

\cite{buder2022galahaccreted} studied accreted stars in GALAH and suggested to use [Na/Fe] instead of [Al/Fe], in order to select accreted stars. In terms of [Na/Fe], part of our sample have sub-solar abundances, while part are [Na/Fe]-enhanced, indicating that part of our sample does have chemical abundances compatible with what is expect classically from accreted stars. But we note that while low abundances of [Na/Fe] and [Al/Fe] might be a strong indicator that a star is accreted, there are overlaps in the higher abundance regime of this elements for \textit{in-situ} and \textit{ex-situ} stars. Therefore, a higher abundance of these elements, does not discard the possibility of a star being accreted, as discussed in \cite{buder2022galahaccreted}. The depletion of [Cu/Fe] might be a chemical indicator that the stars are accreted, since a depletion of this element was reported before for both accreted stars in the Milky Way \citep{nissen2010two,matsuno21,carrillo2022detailed} and in the Small Magellanic Cloud \citep{mucciarelli2023chemical}.

\subsection{Peculiar stars}\label{subsec:peculiar_stars}

In this work, some peculiar stars were identified, and their characteristics warrant further examination. In Section \ref{subsec:dynamics_results}, we found two stars in our sample with low eccentricity. Additionally, in Section \ref{subsection:results_ages} we found two stars with young ages in all the three analysis we performed. In Section \ref{subsec:chemical_abundance_distribution} we found three stars with high [Fe/H], a $\alpha$-poor star, and three Al poor stars.

The two stars with low eccentric orbits have standard low metallicities ([Fe/H] < -0.8) and do not have any peculiar chemical signature when compared with other accreted star candidates in our sample. Therefore we conclude they can be contaminants (stars formed \textit{in situ}) in our sample.

The two stars that have young ages in the seismic analysis are the stars with IDs 407 and 418, and also appears to be young in the other two analysis. Additionally, they appear to have low-metallicities ([Fe/H] < -0.7) in all analysis and high eccentric orbits (\textit{ecc} > 0.7), which agree with the literature of accreted stars. 

The three [Fe/H]-rich stars do not present any remarkable chemical difference compared to other stars in our sample except from being also $\alpha$-rich and have high eccentricity.  These stars may potentially belong to the Splash \citep{belokurov2020biggest} substructure, which is related to the thick disk of the Galaxy. The Splash stars have high eccentric, low angular momentum and in some cases retrograde orbits. 
 They are believed to have formed \textit{in situ} but had their orbits altered by a major merger event, possibly involving GES.

The $\alpha$-poor star found is poor both in Ca and Mg and is the most metal-poor star in our sample. However, this star does not display any other significant chemical peculiarities, possibly because certain key elements such as Al, Ba, and Y could not be measured robustly. This star has high-eccentric orbit and an E and L$_{z}$ compatible with GES stars, suggesting it might still share its origin with them. A detailed follow-up study on this object might shed light on its properties and history.

The Al-poor stars are also poor in Na. Apart from this, those stars do not have any other peculiar chemical signature that might indicate they come from a different progenitor galaxy then the other stars in our sample. They all have high eccentric orbits and E and L$_{z}$ compatible with GES stars.

\section{Summary and conclusion}\label{sec:conclusion}

In this work,  we perform a comprehensive analysis of 30 metal-poor TESS accreted star candidates for which we obtained follow-up \texttt{MIKE} spectra. Given the limited number of metal-poor stars observed by TESS, to improve our understanding on how the use of asteroseismic information impacts the characterization of these stars is crucial. Moreover, carefully characterizing accreted star candidates in terms of chemical abundances, dynamics, and ages is fundamental for improving our understanding of the Milky Way's accretion history and the properties of its various building blocks.

In the literature, different works have associated different regions in the energy (E)-angular momentum ( L$_{z}$) plane to different building blocks of the Milky Way. In order to characterize our sample of stars, we also studied them from a dynamical perspective using \texttt{gala} code. The majority of the stars occupy a region in the E-L$_{z}$ plane that is consistent with GES stars. We identified two stars with very low L$_{z}$ values that could potentially be associated with Sequoia debris. Additionally, we found that 28 of the 30 stars in our sample have high eccentricity orbits, typically with eccentricities exceeding 0.7, supporting their origins as accreted stars.

We also investigated the impact of different surface gravity ($\log g$) and the use of $\nu_\mathrm{max}$ and $\Delta$$\nu$ on age and chemical abundance determinations. For the chemical abundance analysis we adopted three different $\log g$: seismic, predicted and spectroscopic. The mean difference between seismic $\log g$ and predicted $\log g$ is 0.07 dex, with a standard deviation of 0.25 dex. Similarly, the mean difference between seismic $\log g$ and spectroscopic $\log g$ is 0.00 dex, with a standard deviation of 0.30 dex.  In most cases, the differences in $\log g$ values were well within the uncertainties, suggesting satisfactory overall agreement among the different $\log g$ considered.

Next, we explored how the use of the predicted and spectroscopic $\log g$, as well as $\nu_\mathrm{max}$ and $\Delta$$\nu$ impacted in the age determination. We used \texttt{BASTA} code to calculate the stellar ages. Seismic ages (calculated using $\nu_\mathrm{max}$ and $\Delta$$\nu$) yield a median age of $11.3_{-4.1}^{+1.3}$ Gyr, with 15 percent of the sample younger than 7 Gyr. Ages calculated using the predicted $\log g$ result in slightly lower median age of $10.1_{-2.9}^{+1.8}$ Gyr, also with 15 percent of the sample younger than 7 Gyr. Ages obtained using the spectroscopic $\log g$ yield a even lower median age of $9.3_{-4.4}^{+2.0}$ Gyr, with 46 percent of the sample younger than 7 Gyr according to our results. Additionally, statistical uncertainties found for the seismic, predicted and spectroscopic cases are of the order of 10.1, 18.5 and 28.6 percent, respectively. Therefore seismic ages tend to be slightly higher than those obtained using predicted or spectroscopic $\log g$ values and also notably more precise following our results. This study highlights the impact of seismic information on age determination, in particular the potential of \texttt{TESS} data for Galactic Archaeology studies. 

However, a different result was observed when evaluating the impact of different $\log g$ in the chemical abundance determination when using \texttt{iSpec} code. Comparing the chemical abundances calculated using the seismic and predicted $\log g$, we observed that overall the distributions are similar, but differences are observed for calcium (Ca). When comparing chemical abundances using the seismic and spectroscopic $\log g$ we also observed that the distributions were similar except for Ca, and scandium (Sc) abundances. We note that Ca is a chemical element very sensitive to $\log g$.

Finally, we studied the overall chemical distribution of our star sample, and our main findings are listed below:

\begin{itemize}
    \item The sample of stars primarily exhibits high [$\alpha$/Fe] abundances at low metallicities ($\mathrm{[Fe/H]} < -1.0$), consistent with typical halo and thick disk stars. This suggests that if the stars are accreted, the progenitor galaxy had a high star formation rate and the necessary mass to produce enhanced [$\alpha$/Fe] abundances.
    \item Iron-peak elements like Ni, Cr, Co, and Mn show abundance patterns that reflect the influence of Type Ia supernovae (SNIa) and suggest a fast star formation rate with multiple SNIa events enriching the gas.
    \item We identify some stars enhanced in Ba, indicating that neutron-capture enrichment played an important role in the chemical evolution of the environment they were formed.
    \item Odd-Z and light elements like vanadium (V), copper (Cu), scandium (Sc), sodium (Na), and aluminum (Al) exhibit varied abundance patterns, with V and Cu showing increases with [Fe/H]. The accreted stars candidates appear to be poor in Cu when compared with stars likely formed in the Milky Way. Al is generally enhanced in the sample, but three stars appear to be Al-poor.
\end{itemize}

Overall, the chemical abundance patterns of the studied stars provide valuable insights into their origin and the processes that shaped the chemical evolution of their progenitor. These findings contribute to our understanding of the Milky Way's assembly and the role of accretion events in building its stellar populations. We also show the importance of multi-dimensional analyses in unraveling the intricate nature of Galactic stars and the accretion history of the Milky Way.

\section*{Acknowledgements}

We thank the anonymous referee, who provided valuable comments to improve this article. DDBS acknowledges ANID (Beca Doctorado Nacional, Folio 21220843) and Becas UDP for the financial support provided. DDBS also thanks Alexander Ji, Sara Vitali, Diane Feuillet and Henrique Reggiani for the valuable comments. PJ acknowledges Fondo Nacional de Desarrollo Científico y Tecnológico (Fondecyt) Regular Number 1231057. DDBS and PJ acknowledge Millennium Nucleus ERIS NCN2021\_017.

This paper includes data gathered with the 6.5 meter Magellan Telescopes located at Las Campanas Observatory, Chile. We also thank the staff of LCO for making the observation of these stars possible during the Covid-19 pandemic. 

This paper includes data collected by the TESS mission. Funding for the TESS mission is provided by the NASA Explorer Program.

This work presents results from the European Space Agency (ESA) space mission Gaia. Gaia data are being processed by the Gaia Data Processing and Analysis Consortium (DPAC). Funding for the DPAC is provided by national institutions, in particular the institutions participating in the Gaia MultiLateral Agreement (MLA). The Gaia mission website is \url{https://www.cosmos.esa.int/gaia}. The Gaia archive website is \url{https://archives.esac.esa.int/gaia}. This work made use of the Third Data Release of the GALAH Survey (Buder et al. 2021). The GALAH Survey is based on data acquired through the Australian Astronomical Observatory, under programs: A/2013B/13 (The GALAH pilot survey); A/2014A/25, A/2015A/19, A2017A/18 (The GALAH survey phase 1); A2018A/18 (Open clusters with HERMES); A2019A/1 (Hierarchical star formation in Ori OB1); A2019A/15 (The GALAH survey phase 2); A/2015B/19, A/2016A/22, A/2016B/10, A/2017B/16, A/2018B/15 (The HERMES-TESS program); and A/2015A/3, A/2015B/1, A/2015B/19, A/2016A/22, A/2016B/12, A/2017A/14 (The HERMES K2-follow-up program). We acknowledge the traditional owners of the land on which the AAT stands, the Gamilaraay people, and pay our respects to elders past and present. This paper includes data that has been provided by AAO Data Central (datacentral.org.au).

This research also made use of Matplotlib \citep{hunter2007matplotlib}, NumPy \citep{walt2011numpy}, SciPy\citep{jones2001scipy}, Astropy \citep{robitaille2013astropy,price2018astropy} and seaborn \citep{Waskom2021}.

\bibliographystyle{mnras}
\bibliography{bib.bib} 
 
\end{document}